\documentclass[12pt]{article}
\usepackage{epsfig}
\usepackage{amsmath,amsfonts,amsthm}
\usepackage{cite}
\newcommand{\OO}{\mbox{\boldmath $\Omega$}}

\newcommand\ov{\overline}

\newcommand{\ka}{\mbox{\boldmath $\kappa$}}

\newcommand{\rr}{\mbox{\boldmath $r$}}

\newcommand{\p}{\mbox{\boldmath $p$}}

\newcommand{\kk}{\mbox{\boldmath $k$}}

\newcommand\ola{\overleftarrow}
\newcommand\ora{\overrightarrow}

\newcommand\fr{\displaystyle\frac}
\newcommand{\htts}{\mbox{\boldmath$\hat{t}\kern1pt$}}

\begin{document}

\begin{center}
{\Large\bf Investigation of periodic multilayers}\\
V.Bodnarchuck$^1$, L.Cser$^2$, V.Ignatovich$^1$, T.Veres$^2$, S.Yaradaykin$^1$\\
1. FLNP, JINR, Dubna, Russia\\
2. BNC, Budapest, Hungary

\bigskip
\begin{abstract}
Periodic multilayers of various periods were prepared according to
an algorithm proposed by the authors. The reflectivity properties
of these systems were investigated using neutron reflectometry.The
obtained experimental results were compared with the theoretical
expectations. In first approximation, the results proved the main
features of the theoretical predictions. These promising results
initiate further research of such systems.
\end{abstract}
\end{center}

\section{Introduction}

\begin{figure}[b!]
{\par\centering\resizebox*{10cm}{!}{\includegraphics{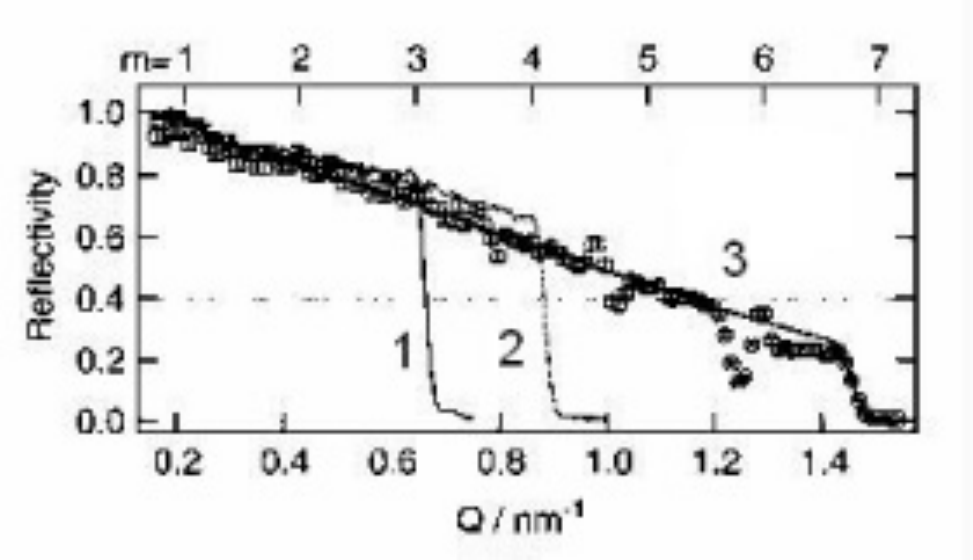}}\par}
\caption{\label{m67} The measured neutron reflectivities of the
Ni-Ti supermirrors fabricated with the large-scale ion-beam
sputtering instrument (IBS) for the Spallation Neutron Source of
J-PARC~\cite{hino3}. Reflectivities of 1) M3 mirror with 403
layers; 2) M4 mirror with 1201 layers; 3) M6.7 mirror with 8001
layers.}
\end{figure}

Neutron supermirrors are nowadays used in many physical
experiments. They are multilayer systems usually composed as a set
of bilayers every one of which consists of two materials with high
and low optical potentials. The supermirrors increase the angular
or wave length range of total reflection comparing to mirrors
consisting of a single material with high optical potential. If
the single material gives total reflection for normal component
$k$ of the incident neutrons in the range $0<k<k_c$, where $k_c$
is the limiting wave number for the given material, the
supermirror can increase the interval up to $nk_c$. A multilayer
system that gives reflectivity $\sim1$ in the interval $0<k<nk_c$
is called Mn mirror. It became a usual practice to fabricate M2
and M3 mirrors. However there are also attempts to produce mirrors
with higher n. The last record belongs to Japanese~\cite{hino3}
who prepared mirror M6.7. Reflectivity of this mirror is shown in
fig.~\ref{m67}.

Last time all the mirrors were prepared in aperiodical fashion,
which means that thicknesses of layers in bilayers vary with the
bilayer number. F. Mezei and P.A. Dagleish performed the first
experimental study of such supermirror in 1977~\cite{2}. The
algorithm for thickness variation was proposed by J.B. Hayter and
H.A. Mook in 1989~\cite{hayt}. According to this algorithm the
change of thicknesses of neighboring bilayers is very small and
does not match interatomic distance. It leads to creation of
unavoidable roughnesses on layers interface. There exists also
another algorithm proposed in~\cite{carron}, according to which
the supermirror is to be produced as a set of periodic chains with
some number N of identical bilayers. The variation of thicknesses
of neighboring chains in this case is larger comparing to
aperiodical systems, which may help to improve the quality of
interfaces and therefore of the whole supermirror.

The goal of the given work is to investigate how well we can
control the thickness and quality of periodic multilayer systems
prepared by Mirrotron Ltd, Budapest. In other words we want to see
how well the neutron reflectivity of produced systems match
theoretical expectations, how large is diffuse scattering because
of technological imperfectness and whether we can explain and
control them.

Below we first present theoretical description of periodic
multilayer systems, and calculate neutron reflectivities of
periodic chains with N bilayers (N=2, 4, 8). Then we present
results of measurements of reflectivities of fabricated
multilayers and compare them to theoretical ones.

\section{Theoretical description of periodical chains}

A periodical chain consists of N bilayers every one of which
contains two layers of different materials. In our case they were
Ni and Ti. We call Ni with higher optical potential a ``barrier'',
and accept the real part of its potential $u_{b0}$, which is
$2.45\cdot10^{-7}$ eV, for unity. It means that the other energies
are defined in units of $u_{b0}$. So, the full Ni potential with
imaginary part is $u_b=1-0.00014i$. We call Ti with lower
potential a ``well'', and its optical potential in units of
$u_{b0}$ is $u_w=-0.203-0.00012i$.

It was decided to investigate periodical stacks, that give Bragg
reflection at the point $k=2$. This point is the normal to the
sample surface component of the incident neutrons wave vector, and
its value is given in units of the critical wave number
$k_c=\sqrt{u_{b0}}$\ of Ni. The point $k=2$ have to be at the
center of the Darwin table of the Bragg peak. Our main task is to
find thicknesses of the two sublayers of a bilayer, to get Bragg
peak (at $N\to\infty$) with maximal width of the Darwin table.

Reflection amplitude of a periodical potential with N symmetrical
periods is given by the equation~\cite{carron}
\begin{equation}\label{a1}
R_N(k)=R_\infty\fr{1-\exp(2iqNa)}{1-R_\infty^2\exp(2iqNa)},
\end{equation}
where
\begin{equation}
R_\infty=\frac{\sqrt{(1+r)^2-t^2}-\sqrt{(1-r)^2-t^2}}{\sqrt{(1+r)^2-t^2}+
\sqrt{(1-r)^2-t^2}}, \label{pe5}
\end{equation}
\begin{equation}
e^{iqa}=\frac{\sqrt{(1+t)^2-r^2}-\sqrt{(1-t)^2-r^2}}
{\sqrt{(1+t)^2-r^2}+ \sqrt{(1-t)^2-r^2}}, \label{pe6}
\end{equation}
and $r$, $t$ are reflection and transmission amplitudes of a
single period.

\begin{figure}[b!]
{\par\centering\resizebox*{6cm}{!}{\includegraphics{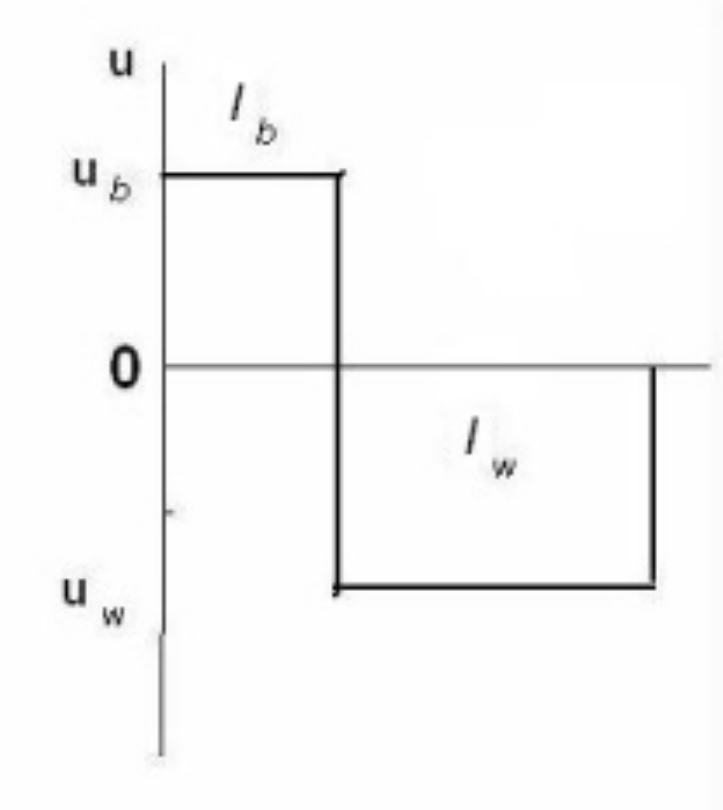}}\par}
\caption{\label{f2} A single element of a multilayered system is a
bilayer composed of two differen materials. The layer of one
material has a high optical potential $u_b$ and a width $l_b$. The
layer of the other material has lower potential $u_w$ and the
width $l_w$.}
\end{figure}

In the case of a bilayer the potential of a period is not
symmetric, as is shown in Fig.~\ref{f2}.
Therefore we have to take into account a direction of reflection
and transmission. We denote $\ora r$ the reflection amplitude for
the incident wave propagating to the right, and $\ola r$ for the
incident wave propagating to the left. Then
\begin{equation}\label{5}
\ora{r}=r_b+t_b^2\fr{r_w}{1-r_br_w},\qquad\ola{r}=r_w+t_w^2\fr{r_b}{1-r_br_w},\qquad
t=\fr{t_bt_w}{1-r_br_w},
\end{equation}
where
\begin{equation}\label{7}
r_{b.w}=r_{0b,w}\fr{1-\exp(2ik_{b,w}l_{b,w})}{1-r_{0b,w}^2\exp(2ik_{b,w}l_{b,w})},\qquad
t_{bw}=\exp(ik_{b,w}l_{b,w})\fr{1-r_{0b,w}^2}{1-r_{0b,w}^2\exp(2ik_{b,w}l_{b,w})},
\end{equation}
\begin{equation}\label{6}
r_{0b,w}=\fr{k-k_{b,w}}{k+k_{b,w}}\qquad
k_{b,w}=\sqrt{k^2-u_{b,w}}.
\end{equation}
We see that the transmission amplitude, $t$, is symmetric, i.e. it
is the same in both directions. We can also introduce the
symmetrized reflection amplitude
\begin{equation}\label{3}
r=\sqrt{\ora{r}\ola{r}},
\end{equation}
then, with account of asymmetry the equation (\ref{a1}) takes the
form
\begin{equation}\label{1a}
\ora{R}_N(k)=\ora{R}_\infty\fr{1-\exp(2iqNa)}{1-R_\infty^2\exp(2iqNa)},
\end{equation}
where $\ora R_\infty$ and $\ora R_N$ inherit the asymmetry of $r$,
i.e.
\begin{equation}\label{3a}
\ora{R}_\infty=\fr{\ora{r}}{r}R_\infty,\qquad
\ora{R}_N=\fr{\ora{r}}{r}R_N,
\end{equation}
and $R_N$, $R_\infty$ are given by (\ref{a1}), (\ref{pe5}) with
symmetrized amplitude (\ref{3}) used for $r$.

To find $l_{b,w}$ of the layers in the bilayer, which at $k=2$
give the center of the widest possible Darwin table with $|\ora
R_\infty|=1$, we first neglect imaginary parts of the potentials
$u_{b,w}$, and represent $t$ in the form $t=|t|\exp(i\psi)$ with
real phase $\psi$. Then $r=\pm i|r|\exp(i\psi)$ with the same
phase $\psi$, and Eq. (\ref{pe5}) can be transformed to
\begin{equation}\label{8}
R_\infty=\fr{\sqrt{\sin\psi(k)+|r(k)|}-\sqrt{\sin\psi(k)-|r(k)|}}
{\sqrt{\sin\psi(k)+|r(k)|}+\sqrt{\sin\psi(k)-|r(k)|}}.
\end{equation}
From it we see that the Bragg reflection takes place when
$\sin^2\psi<|r|^2$, the center of the Darwin table is at
$\sin\psi=0$ and the larger is $|r|$, the wider is the Darwin
table. Therefore we must find the widths $l_{b,w}$ from two
conditions:
\begin{equation}\label{9}
\psi(l_b,l_w,k=2)=\pi,\qquad |r(2)|=\max(|r(l_b,l_w,k=2)|),
\end{equation}
where we had shown dependence of $\psi$ and $|r|$ on widths
$l_{b,w}$. To have the conditional maximum at the point $k=2$ we
are to require maximum at this point of the function
\begin{equation}\label{10}
F(l_b,l_w)=|r(l_b,l_w,k=2)|+\lambda[\psi(l_b,l_w,k=2)-\pi],
\end{equation}
\begin{figure}[b!]
{\par\centering\resizebox*{10cm}{!}{\includegraphics{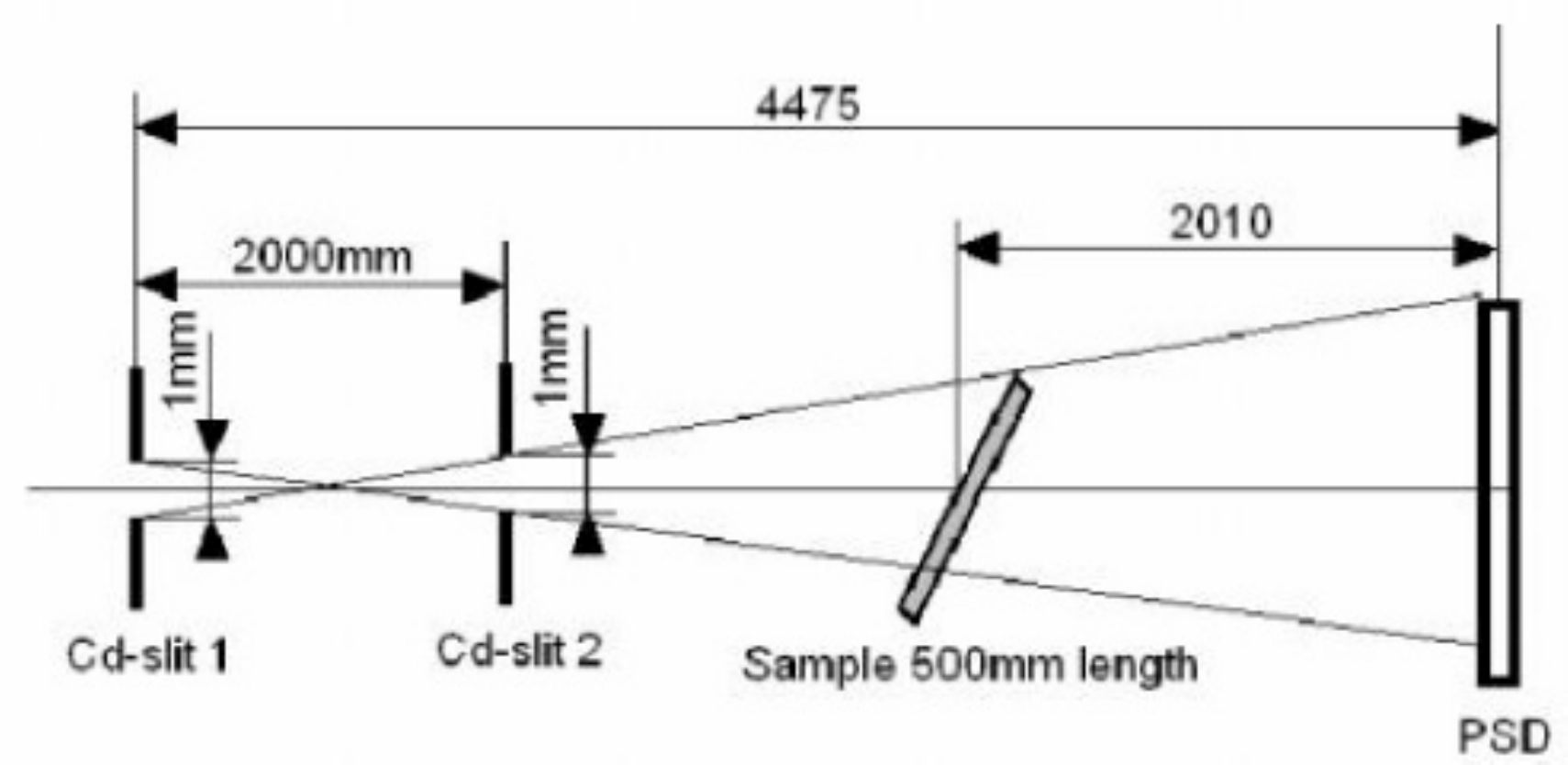}}\par}
\caption{\label{sch} The scheme of the neutron reflectometer at
KFKI. The sample mirror can be rotated around an axis
perpendicular to the plane of the Fig. The position sensitive
detector os stationary and sufficiently large to register all the
reflected neutrons, when grazing angle is sufficiently small
according to experimental requirements. Collimation angle was 0.5
mrad and the neutron beam was monochromatic with wave length 4.28
\AA\ and presumable resolution ? }
\end{figure}
where $\lambda$ is the Lagrange multiplier. The maximum is found
from three equations
\begin{equation}\label{11}
\fr{d}{dl_b},\fr{d}{dl_w},\fr{d}{d\lambda}[|r|-\lambda(\psi-\pi)]=0.
\end{equation}
Solution of these equations gives
\begin{equation}\label{12}
k_bl_b=k_wl_w=\fr\pi2.
\end{equation}
Therefore we must take
\begin{equation}\label{28w}
l_b=\fr{\pi}{2\sqrt{4-u_b}}=0.907,\qquad
l_w=\fr\pi{2\sqrt{4-u_w}}=0.766.
\end{equation}
The unit of length, corresponds to $k_c=1$ (Ni), therefore it is
$\lambda_c/2\pi=92$ \AA. So $l_b=83.4$ \AA, and $l_w=70.4$ \AA. It
was decided to ask preparation of 3 samples with 2, 4 and 8
bilayers with thicknesses: Ni $l_b=84$ \AA, and Ti $l_w=70$ \AA.

\section{Measurement and processing of data}

\begin{figure}[h!]
{\par\centering\resizebox*{16cm}{!}{\includegraphics{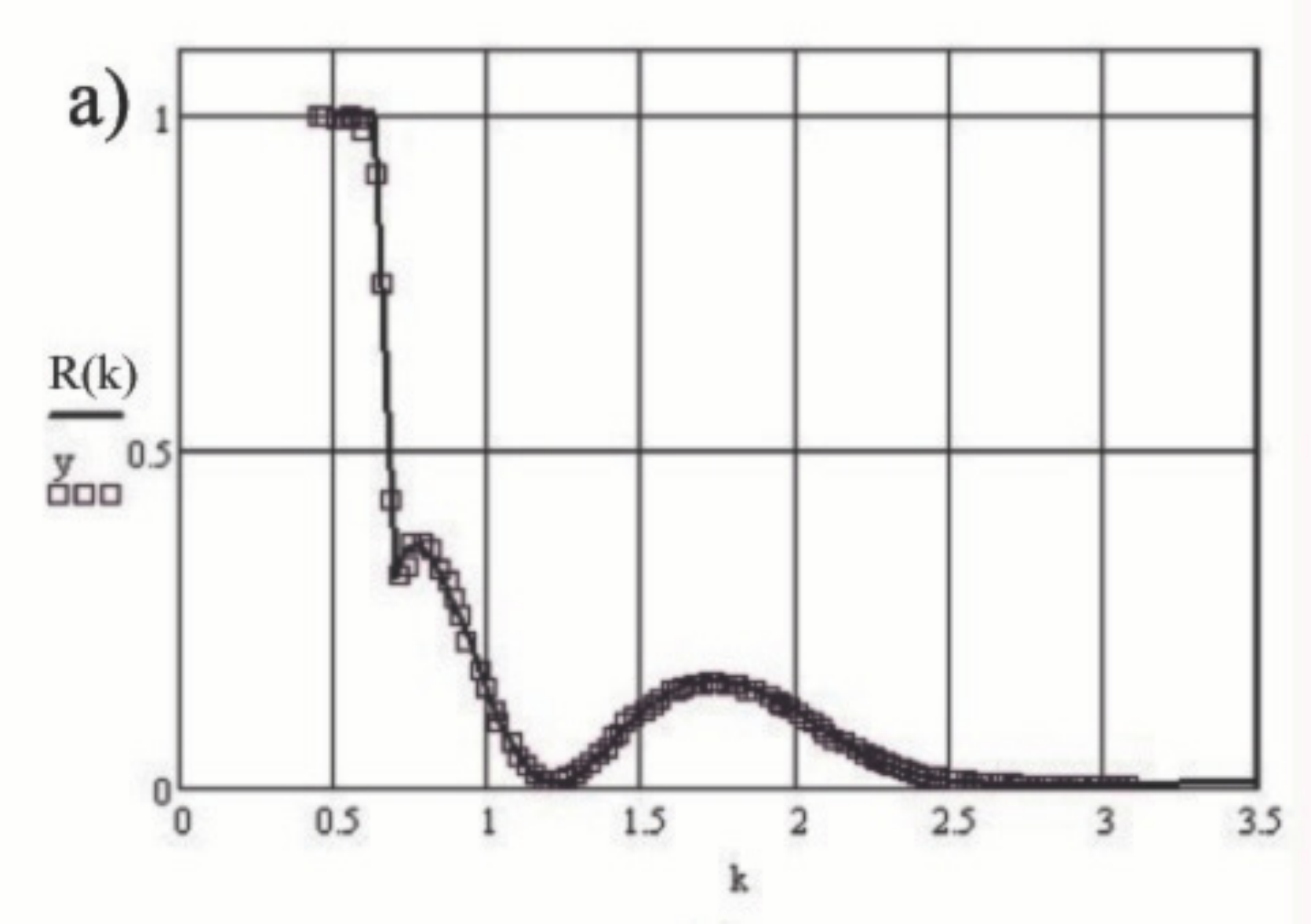}
\hspace{5mm}\includegraphics{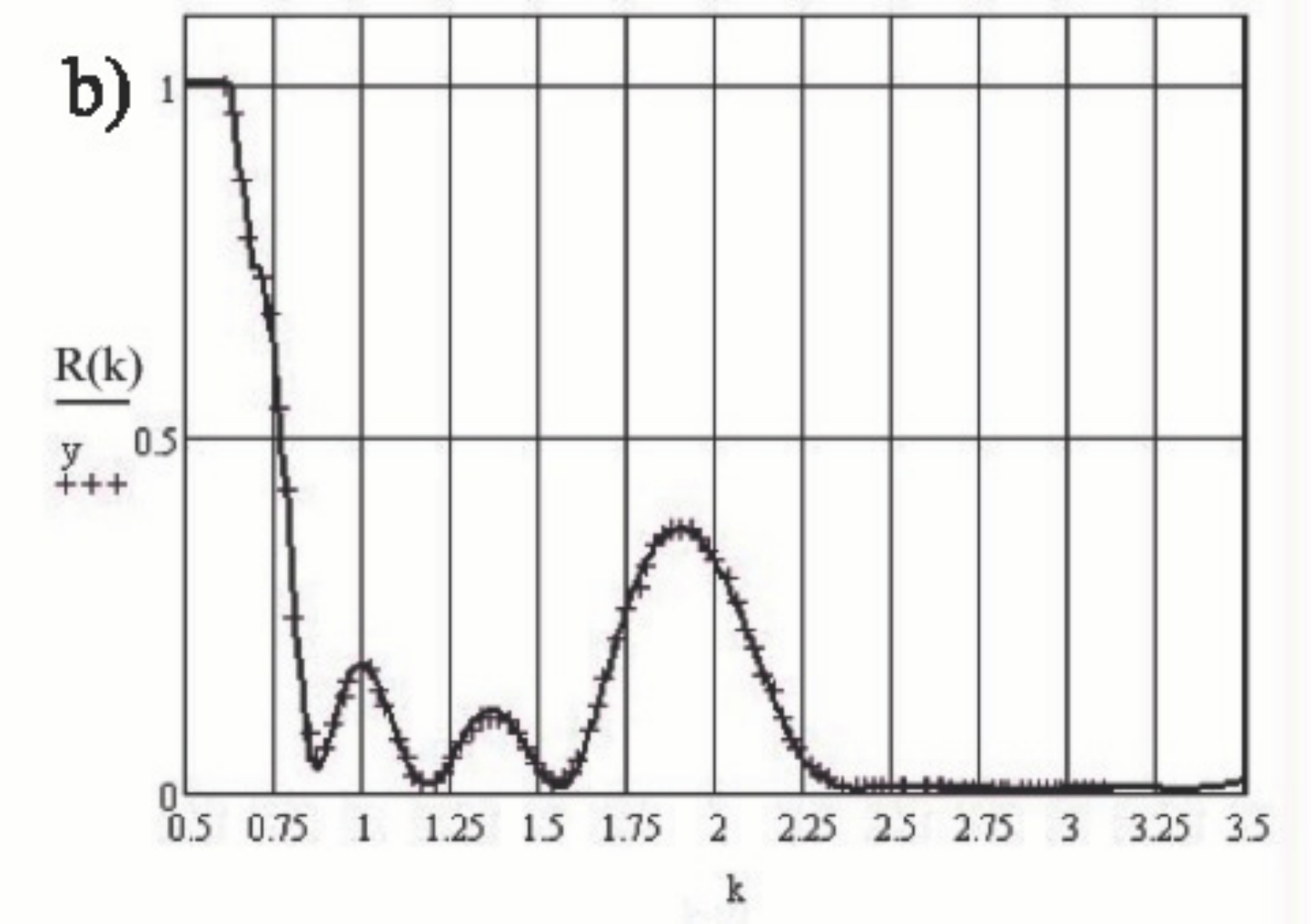}\hspace{5mm}\includegraphics{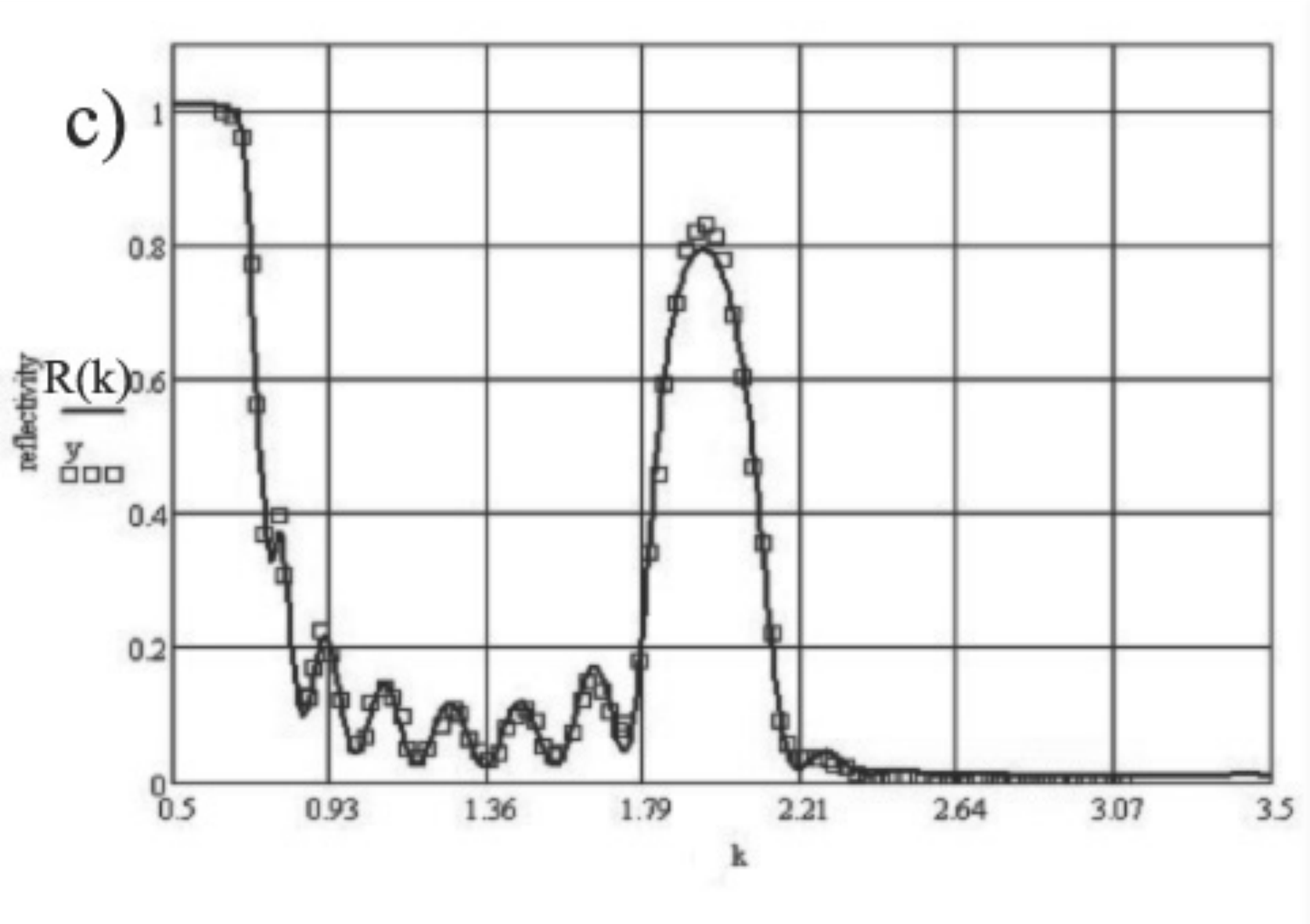}}
\par}
\caption{\label{allp}Fitting of the reflectivity data for periodic
chains of bilayers evaporated on a thick float glass substrate.
The fitting function is given by Equation (\ref{chi3a}). The data
were obtained in Hungary at the reflectometer with wide stationary
detector. The results are shown in linear scale for a) 2; b) 4;
c)8 bilayers. The solid curves are theoretical curves with
parameters found from fitting.}
\end{figure}

In Figure~\ref{allp} in linear scale are shown the results of
fitting experimental data for 2, 4 and 8 periods with the formula:
\begin{equation}\label{chi3a}
R(k)=\int\limits_{k-\delta}^{k+\delta}|\ora{R}_{Ns}(p)|^2\fr{dp}{2\delta}+n_b,
\end{equation}
where
\begin{equation}\label{chi3b}
\ora{R}_{Ns}(p)=\ora{R}_{N}(p)+T^2_N(p)\fr{r_{s0}(p)}{1-\ola{R}_N(p)\,r_{s0}(p)}
\end{equation}
is the reflection amplitude from a periodic chain of N periods
evaporated over a semiinfinite substrate, and the substrate
reflection amplitude is
\begin{equation}\label{rs}
r_{s0}=\fr{k-k_s}{k+k_s},\qquad k_s=\sqrt{k^2-u_s}.
\end{equation}

Eq. (\ref{chi3a}) takes into account the final resolution of the
installation, scheme of which is shown in Fig.~\ref{sch}, and
possible existence of background, $n_b$, in the system. The
resolution $\delta$ and background $n_b$ were two fitting
parameters. Besides them we took as fitting parameters the real
parts of all the potentials $u_b$, $u_w$, the potential of
substrate $u_s$ in terms of that of Ni, and thicknesses of Ni and
Ti layers in units of $\lambda_c/2\pi$ of Ni. Imaginary parts of
the potentials were calculated from absorption cross sections to
be: $u''_b=0.00014$, $u''_w=0.00012$, and $u''_s=0.0001$. The last
one was selected so small, because at first we thought that our
substrate was pure silicon glass.  The results for fitted
parameters are presented in the first three lines of the
Table~\ref{tbl2}. The last column of the table shows $\chi^2$ of
the fitting.
\begin{table}[h!]
\label{tbl2}
\begin{center}
\begin{tabular}{||c|c|c|c|c|c|c|c|c|c|c|c||}
\hline
N&$u'_{b}$ &$u'_{w}$&$u'_{s}$& $u''_{b}$ &$u''_{w}$&$u''_{s}$&$l_b$&$l_w$ &$\delta$&$n_b$&$\chi^2$\\
\hline\hline
2&0.964&-0.258&0.452&0.00014&0.00012&0.0001&1.121&0.59&0.036&0.003&25\\
4&0.934&-0.388&0.446&0.00014&0.00012&0.0001&1.182&0.525&0.033&0.003&114\\
8&0.993&-0.242&0.398&0.00014&0.00012&0.0001&1.061&0.649&0.035&0.0089&349\\
8&0.963&-0.421&0.415&0.00014&0.00012&0.0001&1.169&0.54&0.036&0&209\\
8&0.972&-0.349&0.408&0.00014&0.00012&0.0001&1.13&0.579&0.036&0.003&151\\
\hline
\end{tabular}
\caption{Fitted values of real parts of potentials $u'$ for Ni
(b-barrier), Ti (w-well) and substrate (s), their thickness $l$,
resolution $\delta$, background $n_b$ and $\chi^2$ for periodic
chains with N=2,4,8 bilayers. Imaginary parts $u''$ of the
potentials were fixed. The results are given in dimensionless
units. The unit of energy is equal to real part of Ni optical
potential $u'_{ni}=0.245\,\mu$eV, and unit of length is the
reduced critical wavelength $\lambda_c/2\pi=92$ \AA\ for Ni. The
parameters at the 4-th line were obtained for zero background, and
parameters of 5-th line were obtained for predetermined background
$n_b=0.003$. It is interesting to see that the fitting with
smaller number parameters (6 instead of 7) gave smaller $\chi^2$.}
\end{center}
\end{table}
\begin{figure}[h!]
{\par\centering\resizebox*{16cm}{!}{\includegraphics{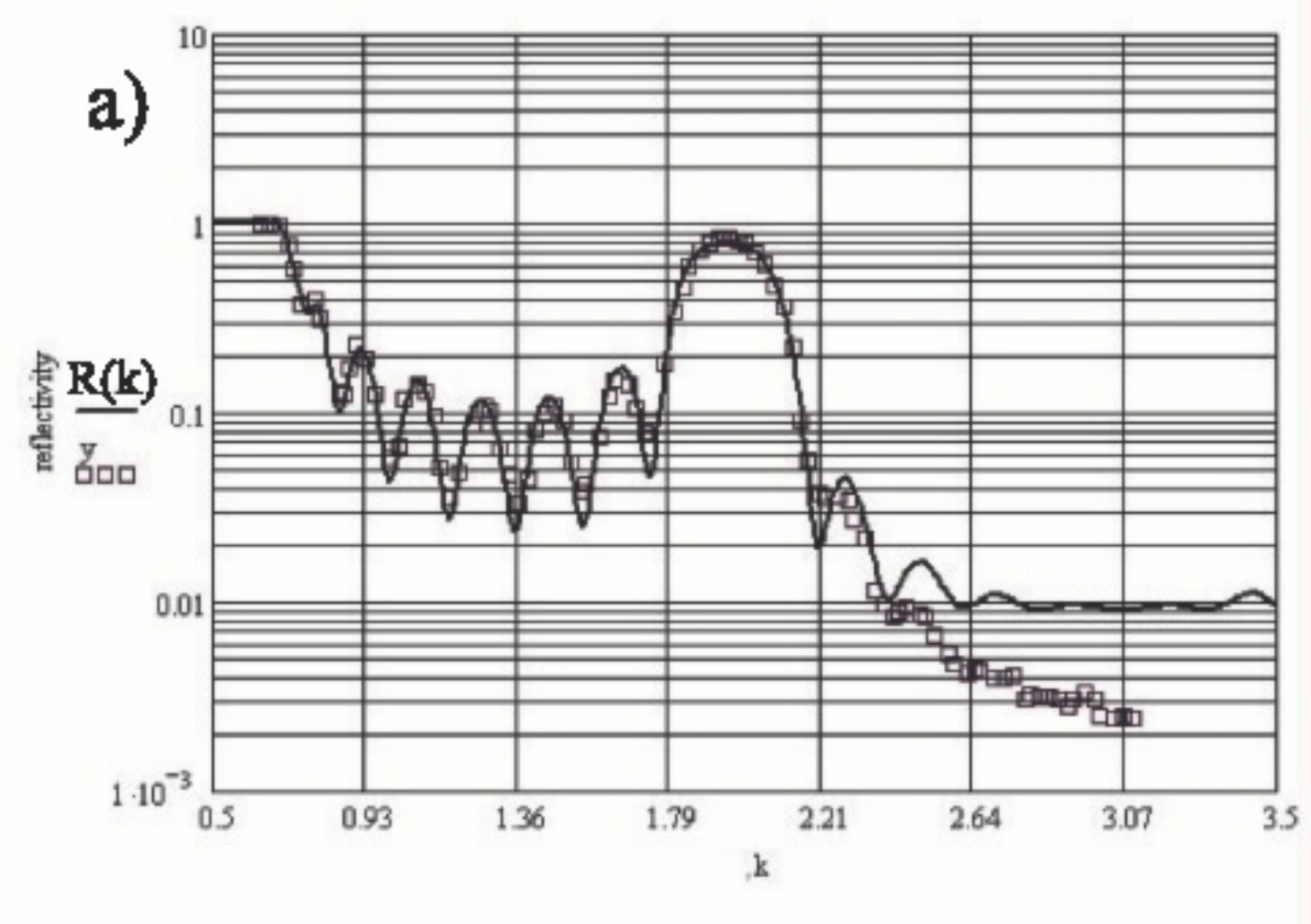}
\hspace{5mm}\includegraphics{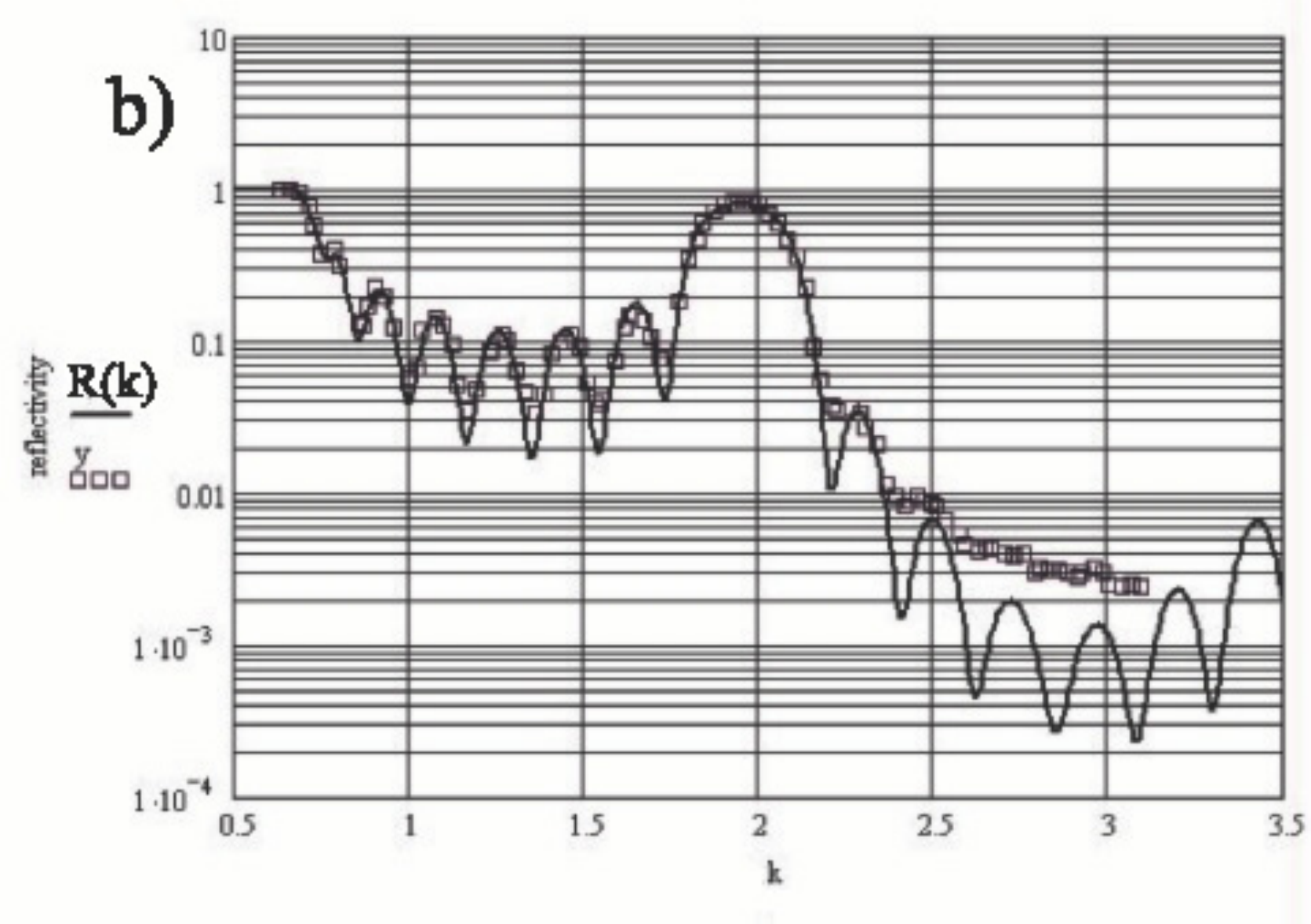}\hspace{5mm}\includegraphics{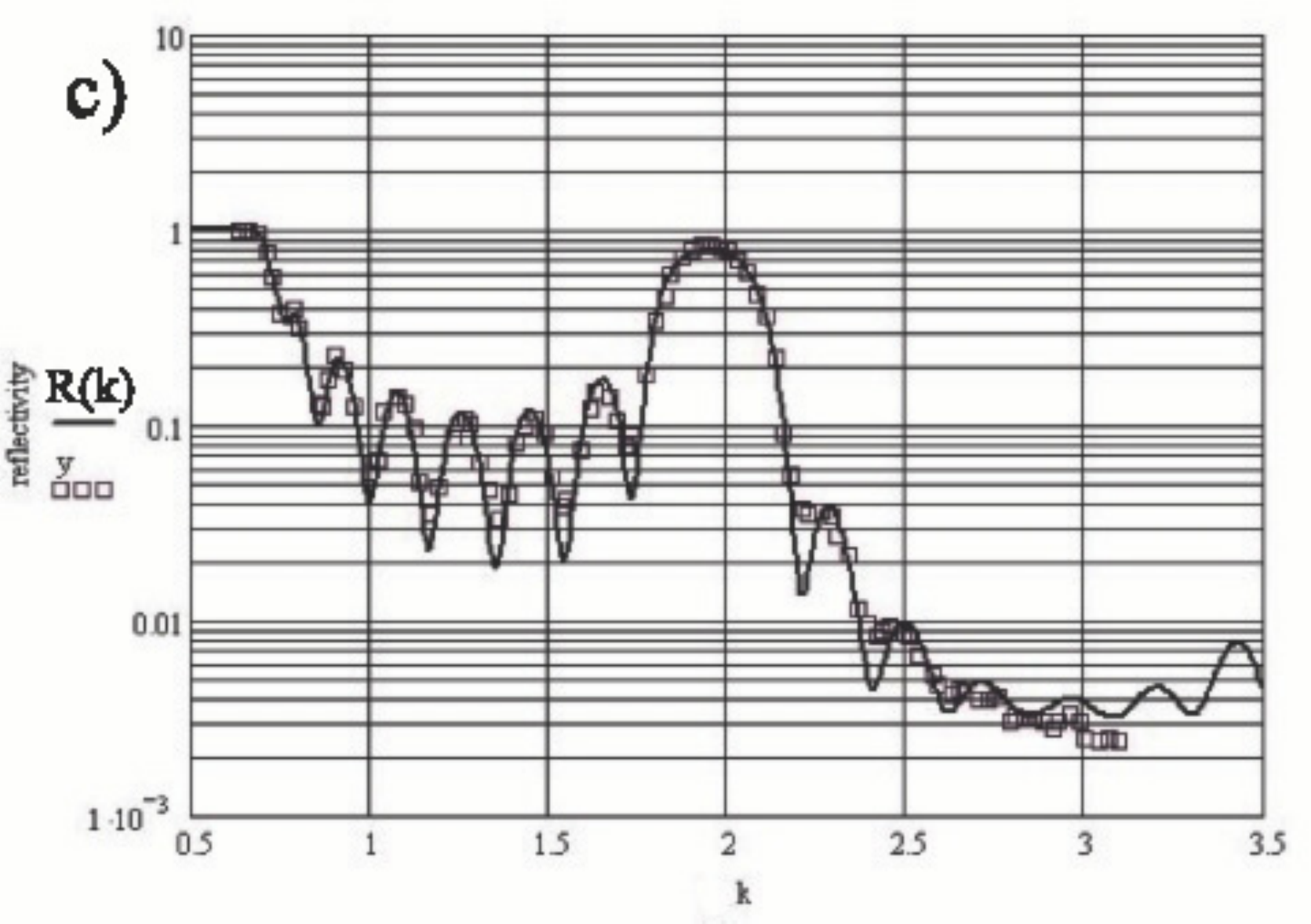}
\hspace{5mm}\includegraphics{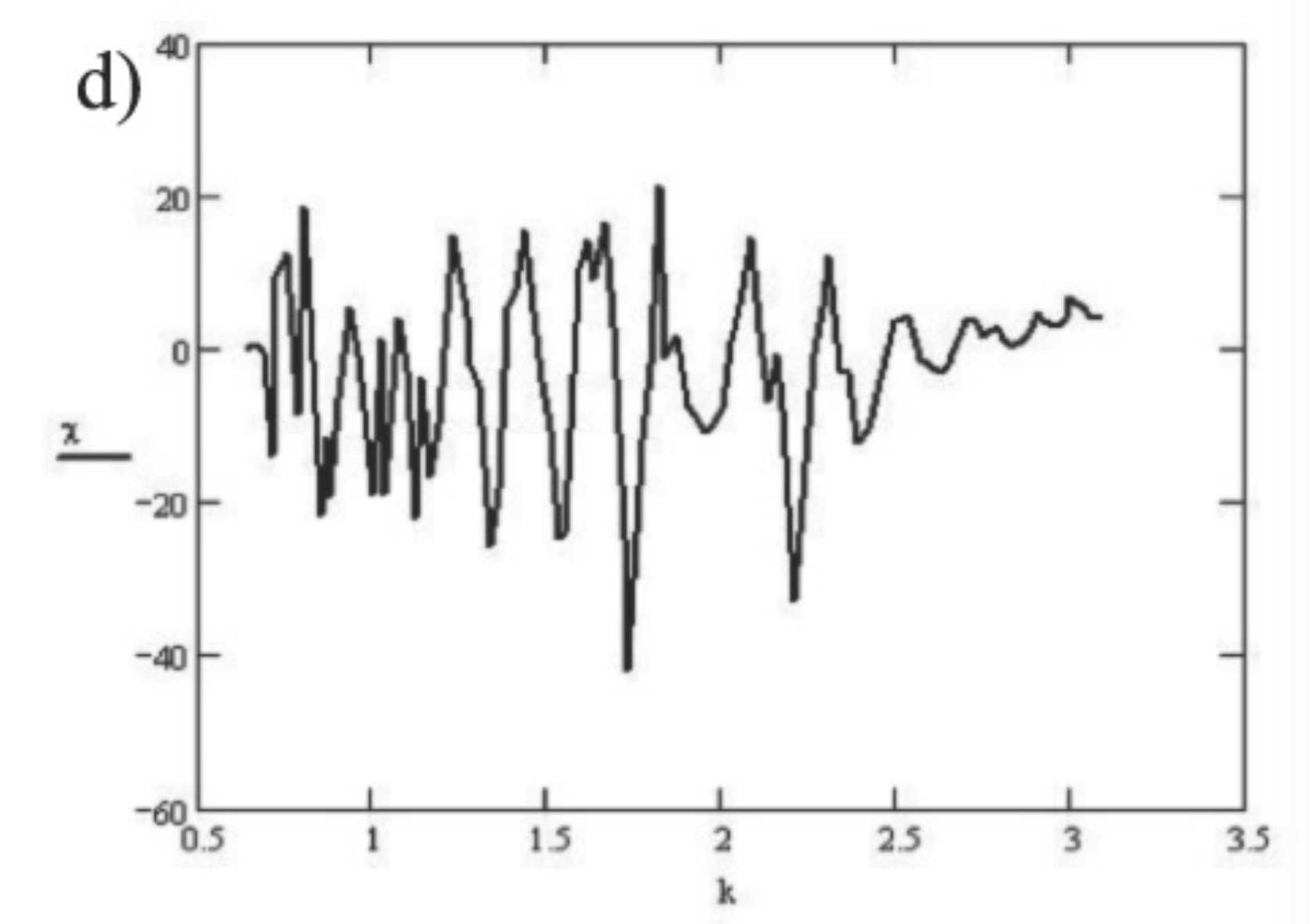}}
\par}
\caption{\label{8bi}Logarithmic scale of different types of
fitting of the reflectivity data from periodic chains of 8
bilayers on a thick float glass substrate. a) The fitting with 7
parameters, corresponding to Figure~\ref{allp}c); b) Fitting with
6 parameters ($n_b=0$); c) Fitting with 6 parameters
($n_b=0.003$); d) $\chi$-distribution in the case c).}
\end{figure}
The pictures in Fig.~\ref{allp} show a good fit of all the
samples, however the thicknesses of the Ni layers are more than
20\% higher and thicknesses of the Ti layers are more than 20\%
lower than in the project. The Ni potential was found to be
slightly lower than is expected, which can be explained by
presence of some oxygen or nitrogen impurities. The substrate
potential was found to be too high comparing to pure silicon
glass, but later we found that it was boron glass the potential
$u_s\approx0.4$ for it is quite reasonable. The resolution
$\delta\approx 3.5$ \% and background $n_b=0.003$ are also quite
acceptable. The worst was the value of the $\chi^2$. It is
especially high in the case of the 8 periods sample.

The defects of fitting of the 8 periods sample are seen in Fig.
~\ref{8bi} in logarithmic scale. In picture a), which corresponds
to picture c) in Fig.~\ref{allp}, it is seen that theoretical line
goes too high at large $k$. It corresponds to overestimated
background $n_b=0.0089$. If we exclude $n_b$ from fitting
parameters and put $n_b=0$ the other fitting parameters are
changed as shown in 4-th line of the Table~\ref{tbl2}. The result
of fitting in logarithmic scale is shown in picture b) of the
Fig.~\ref{8bi}. It is seen that the background in this case is
underestimated. If we put background $n_b=0.003$ fixed at the same
level as was obtained for samples with 2 and 4 periods, we obtain
fitting parameters shown in 5-th line of the Table~\ref{tbl2}, and
result of fitting in logarithmic scale shown in picture c) in
Fig.~\ref{8bi}. The parameter $\chi^2$ decreased (note that the
number of fitting parameters in that case is 6, which is less than
7), however it is still too high, and in Fig.~\ref{8bi}d) there is
presented the $\chi$ distribution
\begin{equation}\label{hik}
\chi(k_j)=\fr{R(k_j)-y(k_j)}{\sigma(k_j)},
\end{equation}
where $y(k_j)$ and $\sigma(k_j)$ are reflectivity and statistical
error at experimentally measured points $k_j)$. This distribution
has very high fluctuations near minima of the data  shown at
picture c) and near the potential edge.

For analysis of the reason of so high fluctuations it was decided
to analyze first the substrate.
\begin{figure}[t]
{\par\centering\resizebox*{16cm}{!}{\includegraphics{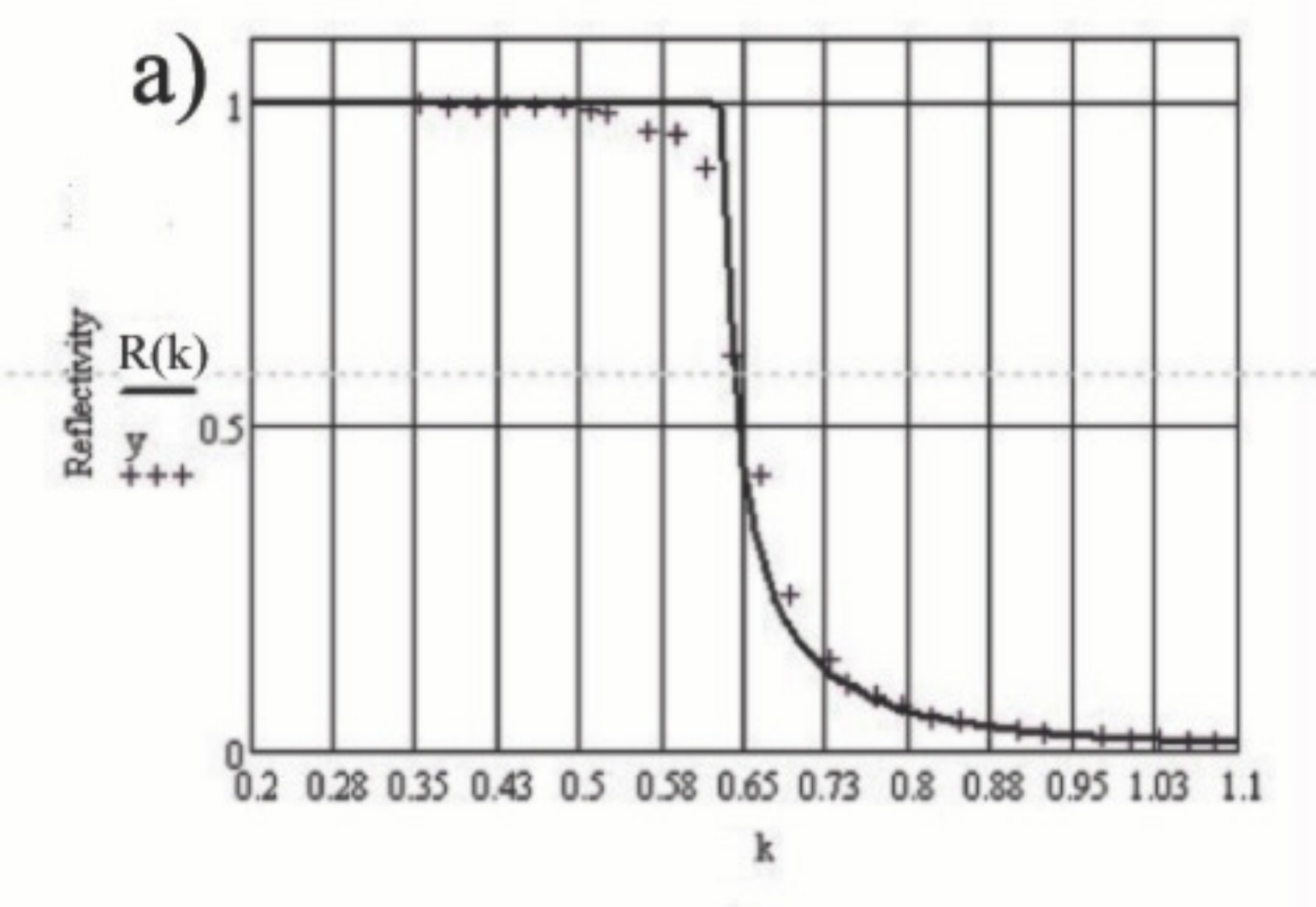}
\hspace{5mm}\includegraphics{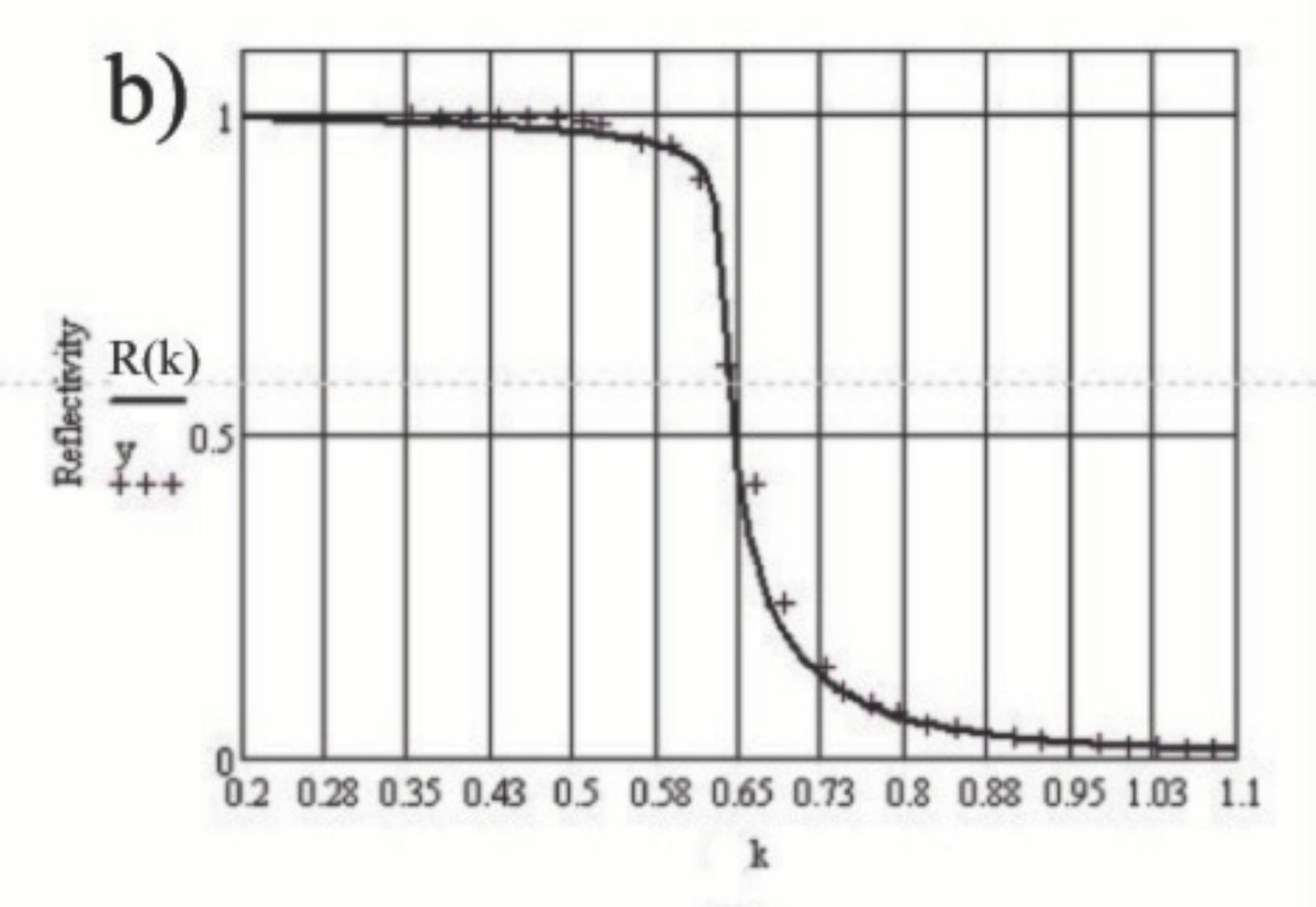}\hspace{5mm}\includegraphics{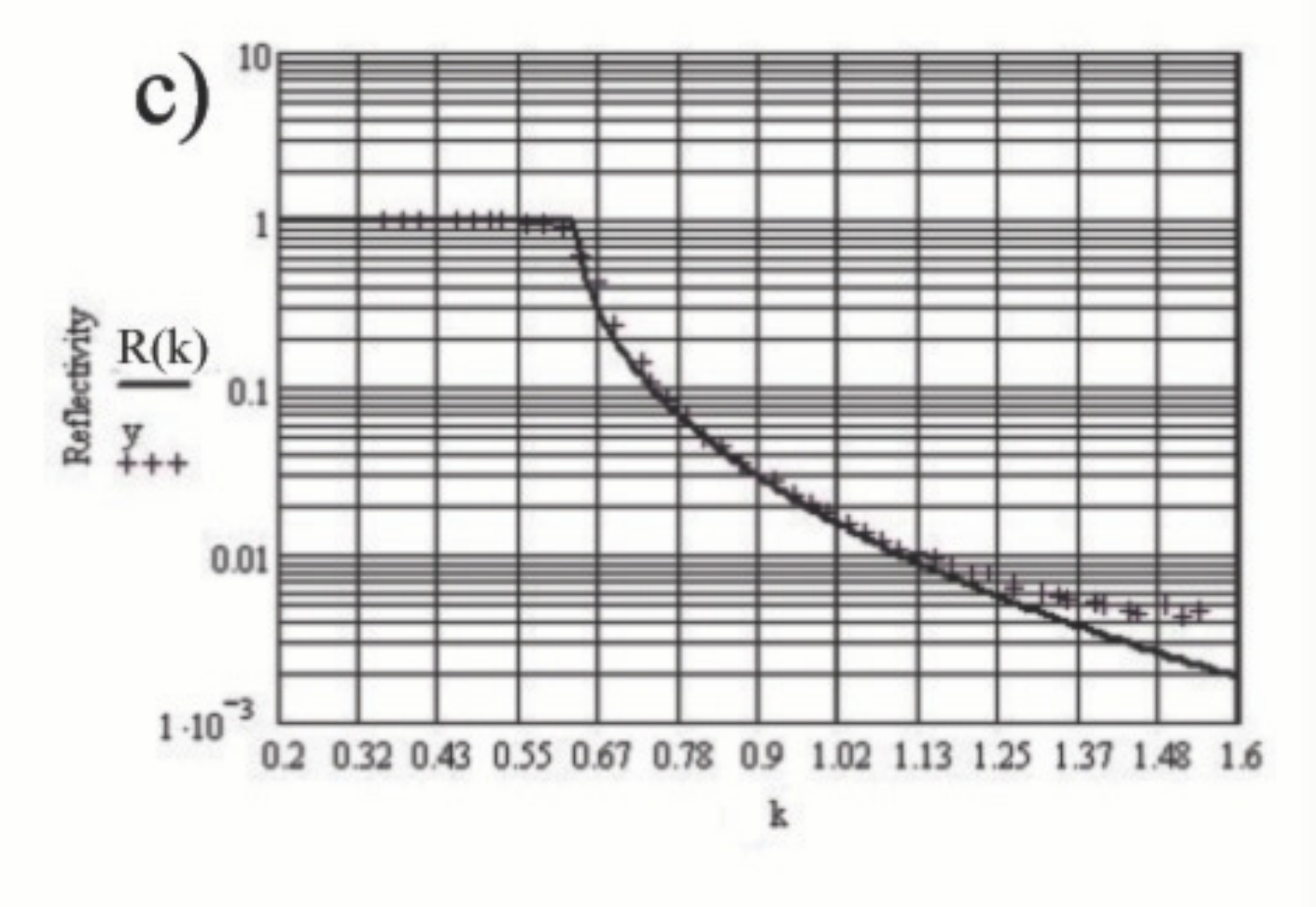}
\hspace{5mm}\includegraphics{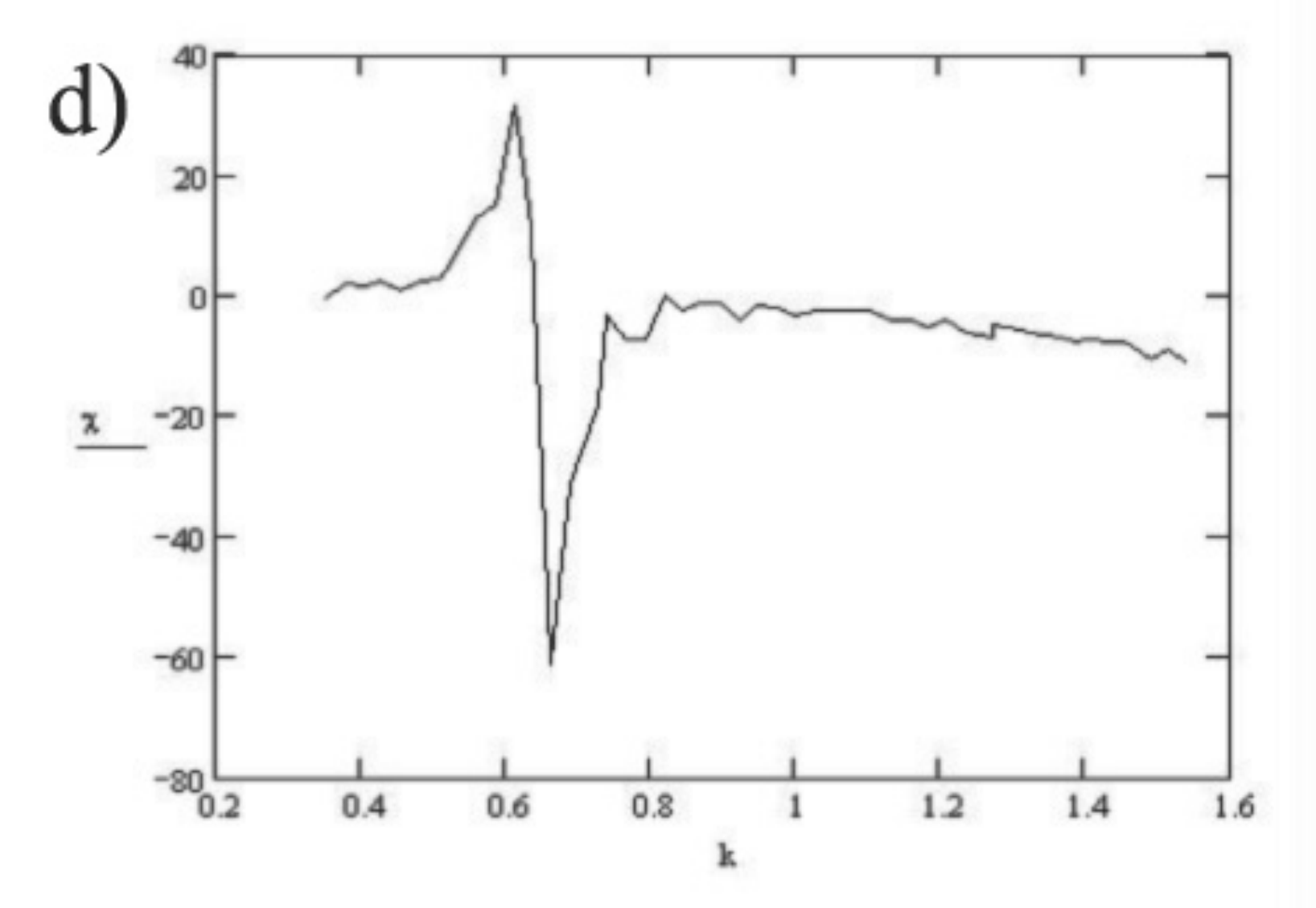}}
\par}
\caption{\label{huglas}Fitting of the glass reflectivity data. a)
One fitting parameter $u'$; b) Two fitting parameters: $u'$ and
$u''$; c) Logarithmic scale of the Figure b); d) $\chi(k_j)$
distribution for fitting with two parameters.}
\end{figure}
\begin{figure}[b]
{\par\centering\resizebox*{16cm}{!}{\includegraphics{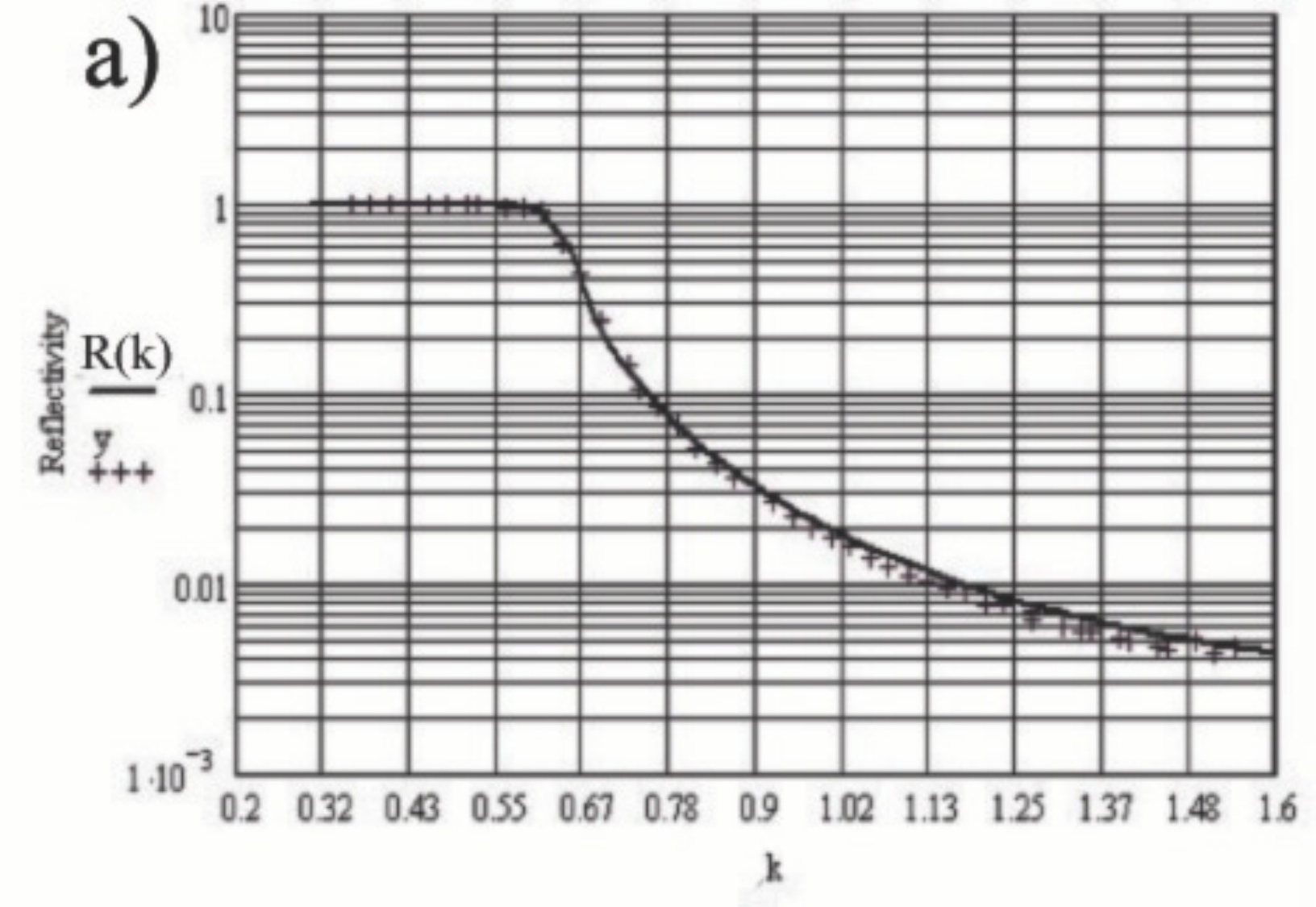}
\hspace{5mm}\includegraphics{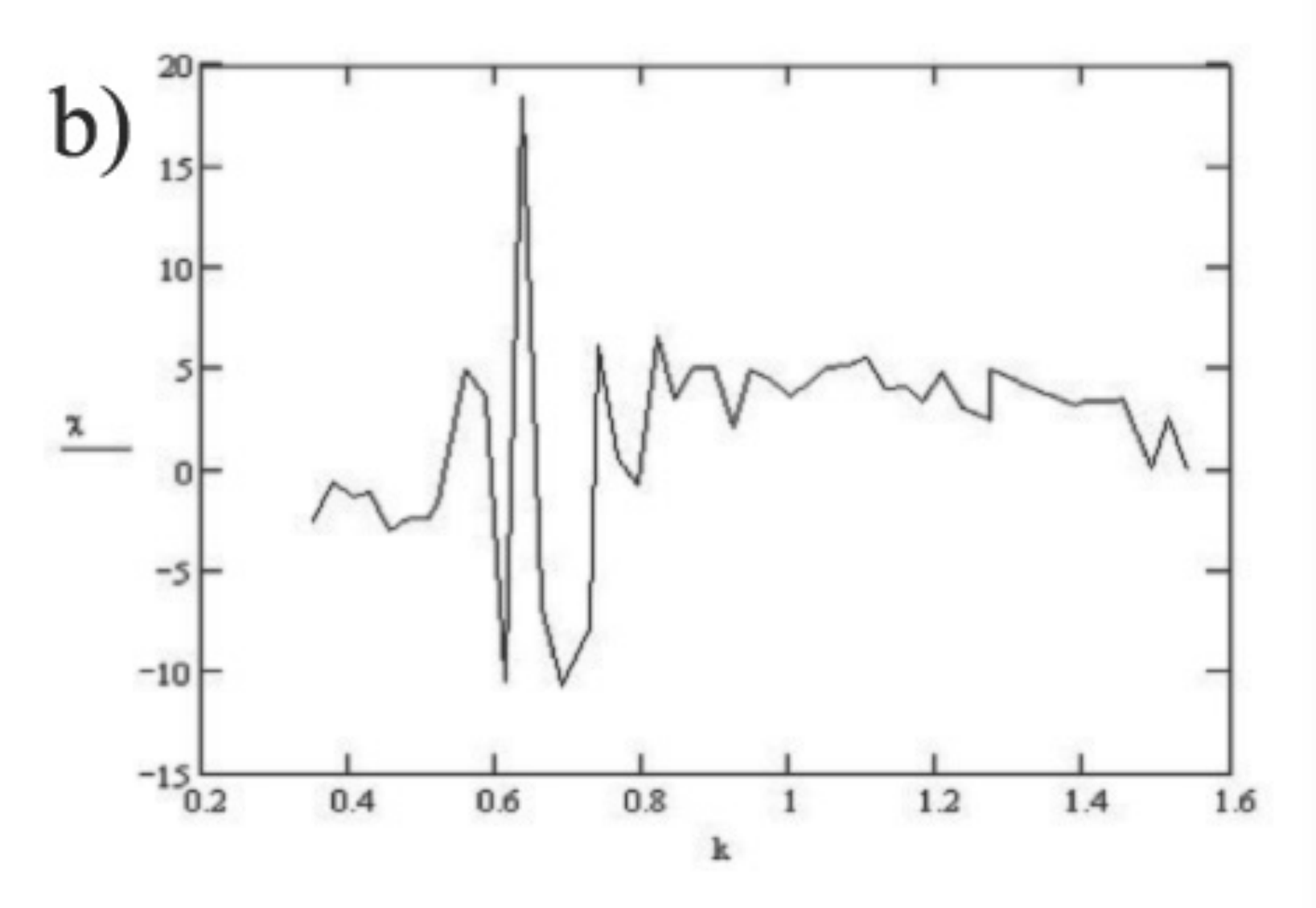}}
\par}
\caption{\label{huglas1}Fitting of the data for glass reflectivity
a) with 4 parameters and $R(k)$ given by (\ref{chi3}); b)
Distribution of $\chi(k)$ for such fitting.}
\end{figure}
In Figure~\ref{huglas} there is presented the fitting of
experimental data for substrate reflectivity fitted by function
\begin{equation}\label{vity}
R(k)=|r_{s0}(k)|^2
\end{equation}
with one fitting parameter $u'$, while $u''=0.0002$ was fixed. The
result was $u'=0.405$, which is in agreement with the fitting of
periodic chains, and proves that the substrate is the boron glass,
however $\chi^2=170$ was too high. In linear scale the graph is
shown in Fig.~~\ref{huglas}a). The fitting with the same function
but with 2 fitting parameters $u'$ and $u''$ gives $u'=0.406$ and
$u''=0.00464$. In linear scale the graph is shown in
Fig.~\ref{huglas}b) and in logarithmic scale in
Figure~\ref{huglas}c). We see a strong deviation at large $k$. The
parameter $\chi^2=137$ for such a fitting becomes a little bit
less but remains still unacceptably high. The distribution
$\chi(k)$ over all the experimental points is shown in
Figure~\ref{huglas}d).

In Fig.~~\ref{huglas1} are shown the result of fitting with more
parameters. The picture a) shows fitting with 4 parameters
according to Equation
\begin{equation}\label{chi3}
R(k)=\int\limits_{k-\delta}^{k+\delta}|\ora{r}_0(p)|^2\fr{dp}{2\delta}+n_b.
\end{equation}
The fitting parameters were $u'$, $u''$, $\delta$ and $u_b$. It
was obtained: $u'=0.408$, $u''=0.003$, $\delta=4$\% and
$n_b=0.0024$, which is quite reasonable. The result of fitting in
logarithmic scale is shown in Figure~\ref{huglas1}a).It looks
quite well, however $\chi^2=31$, for this fitting is still too
high. Distribution $\chi(k)$, shown in Figure~\ref{huglas1}b)
demonstrates that there is still some peculiarity in experimental
data near the critical edge.

\section{Scattering at the interface}

The anomaly near the potential edge of the substrate can appear
because of scattering on surface roughness, on near surface
inhomogeneities and even because of disordered distribution of
atoms inside the glass~\cite{ig}. The scattering at interface can
be calculated with the help of distorted wave Born approximation
(DWBA) i.e. with the help of Green function $G(u,k,\rr,\rr')$,
which takes the interface into account.

\subsection{Green function of DWBA}

The Green function for reflection from the semiinfinite substrate
satisfies the equation
\begin{equation}\label{esr4}
\left(\Delta+k^2-\Theta(z>0)u\right)G(u,k,\rr,\rr')=-4\pi\delta(\rr-\rr'),
\end{equation}
where $u$ is the optical potential of an ideal medium at $z>0$.
Since the space along the interface is uniform, the Green function
can be represented by two dimensional Fourier integral
\begin{equation}\label{esr5}
G(u,k,\rr,\rr')=-\int
\frac{d^2p_\|}{\pi}\exp\left(i\p_\|(\rr_\|-\rr'_\|)\right)G(u,p_\bot,z,z').
\end{equation}
Substitution of it into Eq. (\ref{esr4}) shows that
$G(u,p_\bot,z,z')$ satisfies the equation
\begin{equation}\label{esr4a}
\left(d^2/dz^2+p^2_\bot-\Theta(z>0)u\right)G(u,p_\bot,z,z')=\delta(z-z').
\end{equation}
According to common rules it can be constructed with the help of
two linearly independent solutions $\psi_{1,2}(u,p_\bot,z)$ of the
homogeneous equation
\begin{equation}\label{az9}
\left(d^2/dz^2+p^2_\bot-\Theta(z>0)u\right)\psi_{1,2}(u,p_\bot,z)=0.
\end{equation}
For the function $\psi_1(u,p_\bot,z)$ we can take
\begin{equation}
\psi_1(u,p_\bot,z)=\Theta(z<0)\bigg[\exp(ik_\bot
z)+r_{s0}(k_\bot)\exp(-ik_\bot
z)\bigg]+\Theta(z>0)t_{s0}(k_\bot)\exp(ik'_\bot z), \label{esr2}
\end{equation}
where $r_{s0}(x)$ is reflection amplitude (\ref{rs}), and
$t_{s0}(x)=1+r_{s0}(x)$ is the refraction amplitude from vacuum
into the substrate. For $\psi_2(u,p_\bot,z)$ it is appropriate to
take the function
\begin{equation}\label{esr6}
\psi_2(u,p_\bot,z)=\Theta(z<0) \exp(-ip_\bot
z)t_{0s}(p_\bot)+\Theta(z>0)\left(\exp(-ip'_\bot
z)-r_{s0}(p_\bot)\exp(ip'_\bot z)\right),
\end{equation}
where $t_{0s}(x)=1-r_{s0}(x)$ is the refraction amplitude from
substrate into the vacuum, and $p'_\bot=\sqrt{p^2_\bot-u}$. The
function (\ref{esr6}) contains incident wave propagating inside
the matter. The Wronskian of the functions
$\psi_{1,2}(u,p_\bot,z)$ is $w_{12}(p_\bot)=2ip_\bot
t_{0s}(p_\bot)$.

With functions $\psi_1(u,p_\bot,z)$ from (\ref{esr2}) and
$\psi_2(u,p_\bot,z)$ from (\ref{esr6}) we have
\begin{equation}\label{aza6}
G(u,p_\bot,z,z')=\frac{\left(\Theta(z>z')\psi_1(u,p_\bot,z)\psi_2(u,p_\bot,z')+
\Theta(z'>z)\psi_1(u,p_\bot,z')\psi_2(u,p_\bot,z)\right)}{2ip_\bot
\tau'(p_\bot)}.
\end{equation}
\subsection{Scattering because of disorder\label{diso}}

We first consider scattering because of disorder and incoherent
scattering. Every atom $j$ inside the ordered medium composed of
coherent scatterers is enlightened by the coherent wave field
$\varphi(\rr_j)=t_{s0}(k_\bot)\exp(i\kk\rr_j)$ created by the
incident wave $\exp(i\kk\rr)$. However in presence of incoherency
and disorder the enlightening field has fluctuations, which we
denote $\delta\varphi(\rr_j)$, and suppose that their average
$\ov{\delta\varphi(\rr_j)}=0$. These fluctuations produce
scattered field
\begin{equation}\label{8ncor}
\delta\Psi(\rr)=\sum\limits_j
G(u,k,\rr,\rr_j)\delta\varphi(\rr_j)b_j=\int
d^2p_\|\exp(i\p\rr)A(\kk,\p),
\end{equation}
where
\begin{equation}\label{rs1}
A(\kk,\p)=\frac{i\ora{t_0}(p_\bot)}{2\pi p_\bot}
\sum\limits_j\exp(-i\p'\rr_j)\delta\varphi(\rr_j)b_j,
\end{equation}
the wave vector $\p$ for the wave scattered into vacuum at $z<0$
is $\p=(\p_\|,-p_\bot)$, and $\p_\bot=\sqrt{k^2-p_\|^2}$. We can
suggest that scattering amplitudes $b_j$ of $j$-th atom, and its
fluctuating field are not correlated, therefore
$\ov{\delta\varphi(\rr_j)b_j}=0$, and the correlation of this
product for different atoms is described by the correlation
function
$$\ov{\delta\varphi(\rr_j)b_j\delta\varphi(\rr_l)b_l|}=K\ov{|b|^2}|\varphi|^2(\rr_j)\{\delta_{jl}+g(|\rr_j-\rr_l|)\},$$
i.e. it is naturally supposed that correlation is proportional to
$|\varphi|^2$ itself.

With the wave function (\ref{8ncor}) we can find the flux of
scattered neutrons in the vacuum through any plane parallel to the
substrate interface
\begin{equation}\label{mrrv1}
J_\bot=\int\limits_S
\frac{d^2r_\|}{2i}\left\langle\left[\ov{\delta}\Psi^*(\rr)\frac{d}{dz}\ov{\delta}\Psi(\rr)-
\ov{\delta}\Psi(\rr)\frac{d}{dz}\ov{\delta}\Psi^*(\rr)\right]\right\rangle=(2\pi)^2
\int\limits_S p_\bot \langle|A(\kk,\p)|^2\rangle d^2p_\|,
\end{equation}
where $*$ means complex conjugate, and $S$ is some large area of
the plane, over which we integrate. Ratio of this flux to the
incident one $Sk_\bot$ gives scattering probability
\begin{equation}\label{mrrvv2}
w(\kk)=\frac{J_\bot(\kk)}{Sk_{\bot}}= \fr{(2\pi)^2}{S} \int
\fr{p_\bot}{k_\bot} \langle|A(\kk,\p)|^2\rangle d^2p_\|.
\end{equation}
Since for scattered waves $\p^2=k^2$, and for propagating waves
$p_\bot=\sqrt{k^2-p_\|^2}$ is a real number, then
\begin{equation}\label{corf2}
d^2p_\|/p_\bot=2d^3p\delta(p^2-k^2)=2\pi dp_\bot,
\end{equation}
and the last equality is correct when we can integrate over angle
$d\phi$ around normal. As a result (\ref{mrrvv2}) is transformed
to
\begin{equation}\label{mrrv2}
w(\kk)=\fr{(2\pi)^3}{S}\int\limits_{0}^k \frac{p_\bot^2}{k_\bot}
\ov{|A(\kk,\p)|^2}dp_\bot.
\end{equation}
With it we can define differential probability, or indicatrix
\begin{equation}\label{corsp1a}
w(k_\bot\to
p_\bot)=(2\pi)^3\fr{p^2_\bot}{Sk_\bot}\ov{|A(\kk,\p)|^2},
\end{equation}
where
\begin{equation}\label{ncor8}
\ov{|A(\kk,\p)|^2}=\left|\frac{t_{0s}(p_\bot)}{2\pi
p_\bot}\right|^2\ov{\sum\limits_{j,j'}e^{-i\ka'^*\rr_j}e^{i\ka'\rr_{j'}}\delta\varphi(\rr_j)b_j
\delta\varphi(\rr_{j'})b_{j'}},
\end{equation}
where $\ka'=(\kk_\|-\p_\|,k'_\bot+p'_\bot)$. The double sum has
diagonal part, which after transformation to the integral over
$N_0d^3r_j$ gives
\begin{equation}\label{ncor8d}
\ov{|A(\kk,\p)|^2}_d=\fr{N_0
\ov{|b|^2}}{2(k''_\bot+p''_\bot)}\left|\frac{t_{s0}(p_\bot)t_{s0}(k_\bot)}{2\pi
p_\bot}\right|^2S,
\end{equation}
and indicatrix
\begin{equation}\label{ncor8d1}
w_d(k_\bot\to p_\bot)=N_0
\ov{|b|^2}\fr{|t_{s0}(p_\bot)t_{s0}(k_\bot)|^2}{2k_\bot(k''_\bot+p''_\bot)}.
\end{equation}
\begin{figure}[!b]
{\par\centering\resizebox*{16cm}{!}{\includegraphics{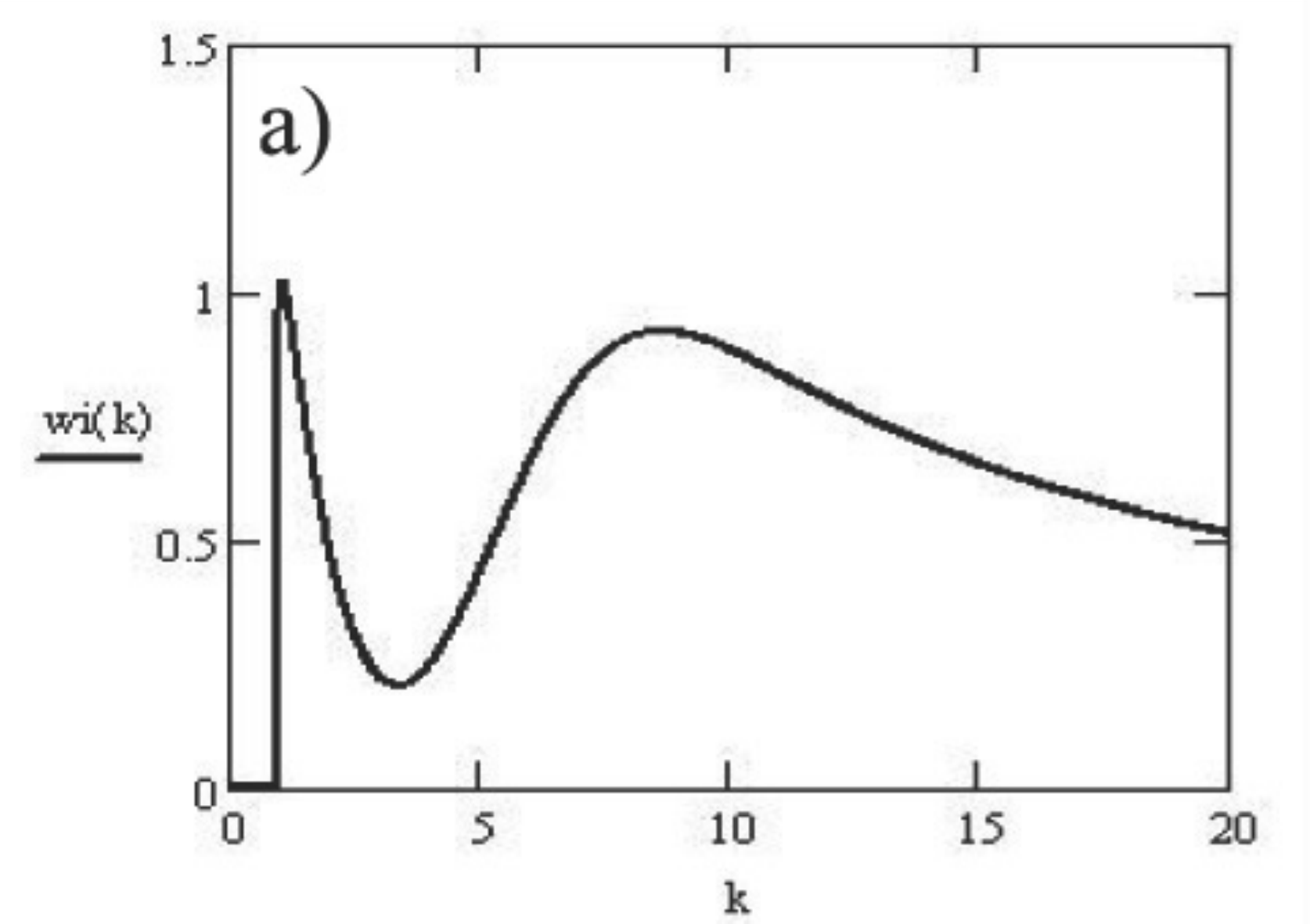}\hspace{15mm}
\includegraphics{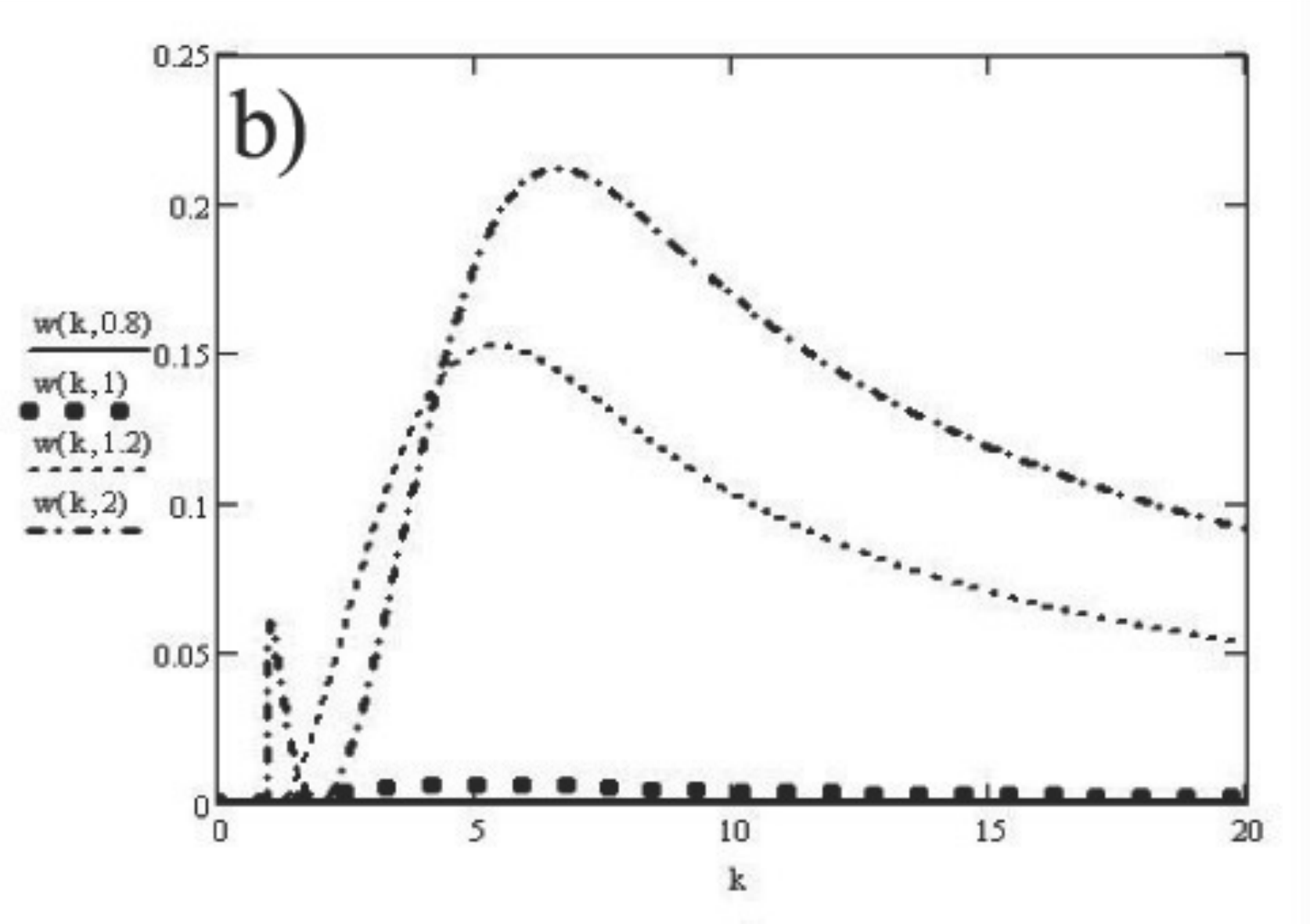}\hspace{15mm}\includegraphics{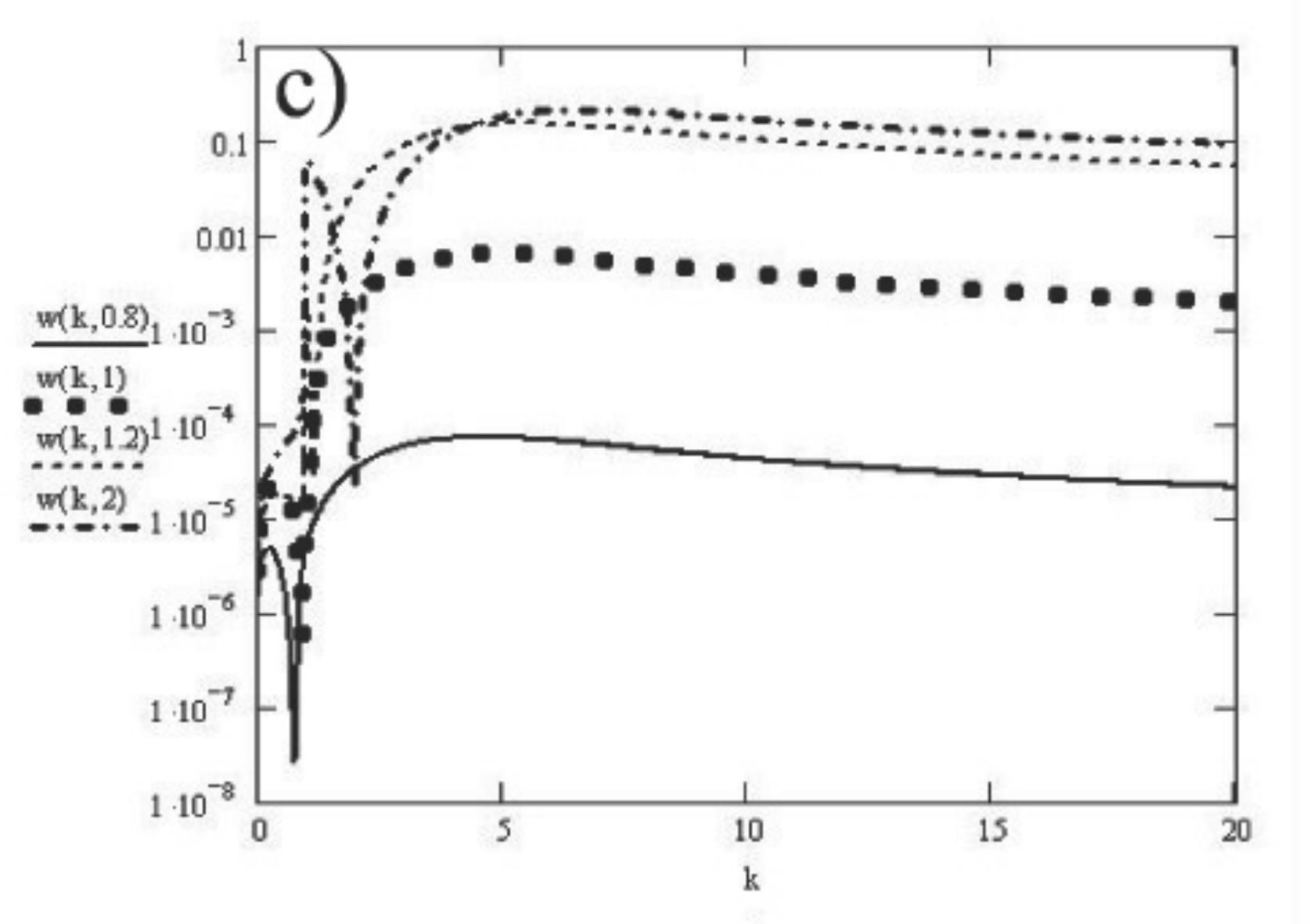}}\par}
\caption{\label{wig}a) Function (\ref{chi83}) in dependence on $k$
for $c=0.0001$, $u=1-i0.0002$, $a^2=0.2$, $b=5$; b) Function
(\ref{m8m4a1}) in linear scale for several different $p$; c) the
same as b) in logarithmic scale.}
\end{figure}
The nondiagonal part of the sum
 can be transformed to double integral:
\begin{equation}\label{dotr}
\ov{\sum\limits_{j,j'}F^*(\rr_j)F(\rr_{j'})}=N_0\int d^3r_j\int
g(\rr_j-\rr_{j'})F^*(\rr_j)F(\rr_{j'}),
\end{equation}
where
\begin{equation}\label{cofa}
g(\rr)=\delta(\rr)-N_0\gamma(\rr)
\end{equation}
is correlation function chosen in such a way as to exclude the
diagonal part of the sum. For simplicity we shall take for
$\gamma(r)$ the Gaussian
\begin{equation}\label{gamm}
\gamma(\rr)=(2\pi)^{-3/2}\exp(-r^2/2a^2),\quad a=N_0^{-1/3}.
\end{equation}
After substitution of $\phi(\rr)$ into (\ref{ncor8}) and
integration we obtain the expression
$$w_{nd}(\OO_0\to\OO)=|b_c|^2 N_0\frac{1}{k_\bot}|t_{s0}(p_\bot)t_{s0}(k_\bot)|^2
\frac{1-\exp(-a^2\ka^2/2)}{p''_\bot+k''_\bot}\approx$$
\begin{equation}\label{mr8wm4a}
\approx\frac{c}{k_\bot}|t_{s0}(p_\bot)t_{s0}(k_\bot)|^2
\frac{1-\exp(-a^2[p'_\bot-k'_\bot]^2/2)}{p''_\bot+k''_\bot},
\end{equation}
where in the last equality we put $c=|b_c|^2N_0$, and approximated
$\kappa^2\approx[p'_\bot-k'_\bot]^2$, which is valid for small
$p'_\bot$ and $k'_\bot$ and small angle between $\kk$ and $\p$.

The function (\ref{mr8wm4a}) represented in the form
\begin{equation}\label{m8m4a1}
w_{nd}(k,p)=\frac{c}{k}\frac{\left|t_{s0}(p)t_{s0}(k)\right|^2}{p''+k''}[1-\exp(-a^2[p'-k']^2/2)],
\end{equation}
and its integral,
\begin{equation}\label{chi83}
wi(k)=c\int\limits_0^b\frac{dp}{k}\frac{\left|t_{s0}(p)t_{s0}(k)\right|^2}{p''+k''}
[1-\exp(-a^2[p'-k']^2/2)],
\end{equation}
for $c=10^{-4}$, $a^2=0.2$, $b=5$ are shown in Figure~\ref{wig}.
The curves are sensitive to parameter $a$ and the upper
integration limit $b$. With increase of $b$ the left maximum in
Figure~\ref{wig}a) increases comparing to the right one, and the
right peak shifts to higher $k$.

\subsection{Fitting of the experimental data for boron glass}

The function (\ref{chi83}) has two maxima, so we can hope to get
better fitting for glass reflectivity shown in
Figure~\ref{huglas1}a). In fitting we accepted potential of Boron
glass to be $u=0.408-0.003i$ as was obtained above. We also
included smoothing of the interface with Debye-Waller factor, i.e
instead of $\ora{r_0}(k)$ we used
$r_m(k)\equiv\ora{r_0}(k)\exp(-2h^2kk')$, where $2h^2$ was a
fitting parameter.

\begin{figure}[!b]
{\par\centering\resizebox*{16cm}{!}{\includegraphics{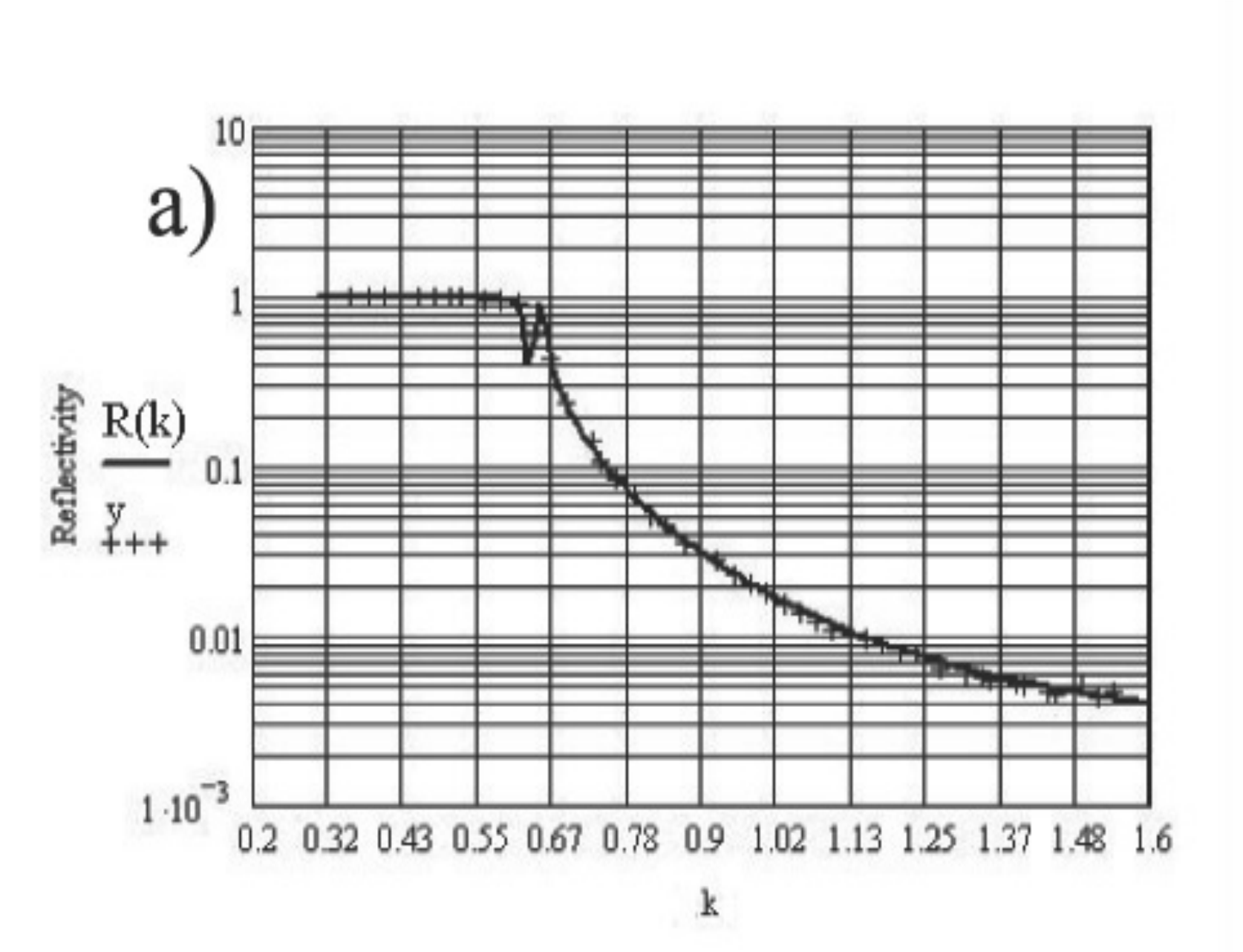}\hspace{25mm}
\includegraphics{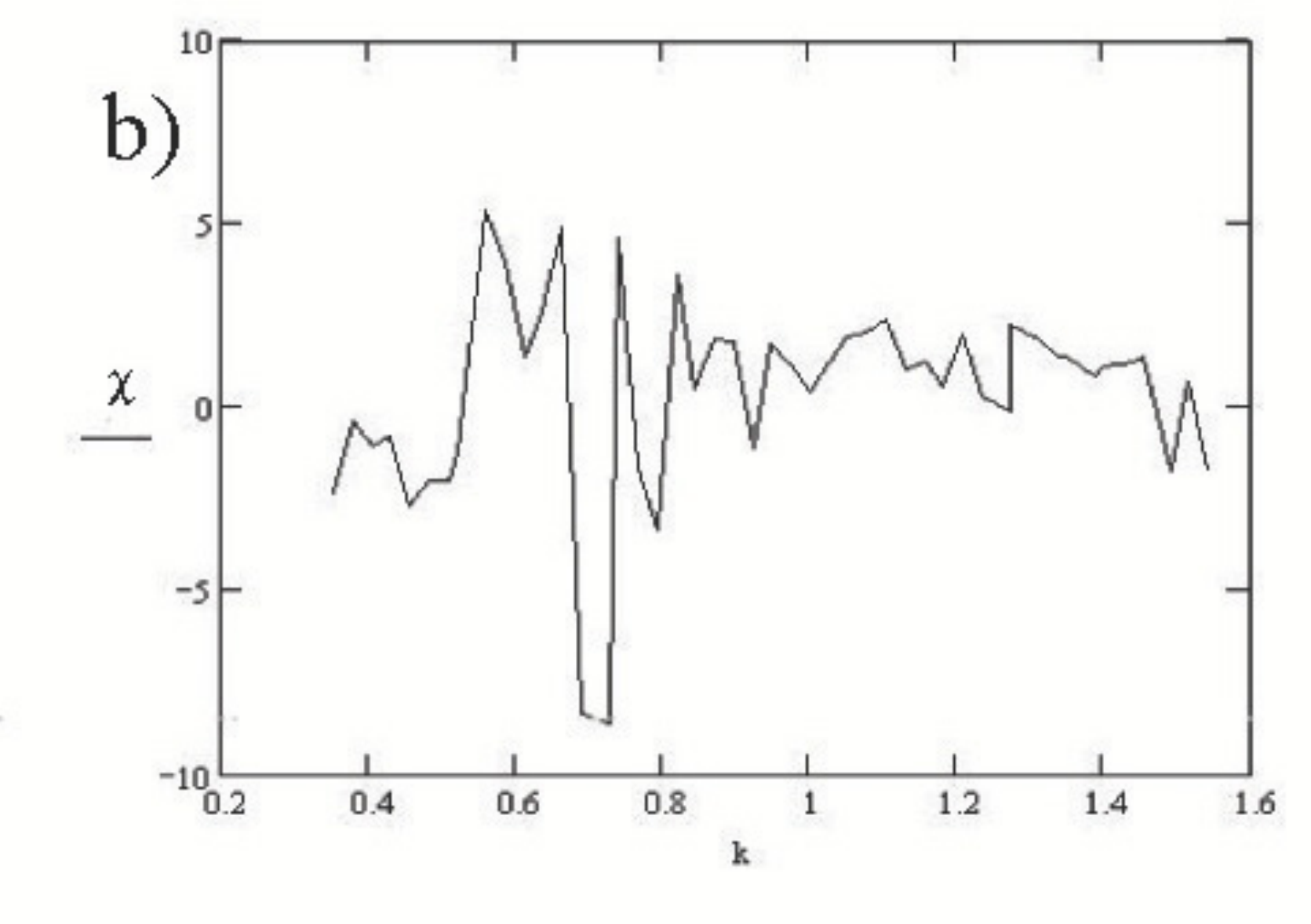}}\par}
\caption{\label{fit5}a) Result of fitting in logarithmic scale of
reflectivity from boron glass with function (\ref{chi883}). The
glass potential $u=0.408-0.003i$ was fixed. The fitting parameters
were found to be: $c=6.323\cdot10^{-4}$, $a^2/2=0.102$,
$\delta=0.022$, $n_b=2.401\cdot10^{-3}$ and $h^2=0.048$. All the
values are quite reasonable. b) $\chi$ distribution for such
fitting. We see that some fluctuations still remain near the edge,
but their amplitude decreased much comparing to that of
Figure~\ref{huglas1}d).}
\end{figure}
Scattering on randomness and fluctuations was represented by
function
\begin{equation}\label{chi883}
w_1(k)=c\int\limits_0^b\frac{dp}{k}\frac{\left|t_{s0}(p)t_{s0}(k)\right|^2}{p''+k''}[2-\exp(-a^2[p'-k']^2/2)].
\end{equation}
The number 2 in brackets shows that we take the sum of two
functions (\ref{mr8wm4a}) and (\ref{ncor8d1}). In the function
(\ref{chi883}) there are two fitting parameters: factor $c$ and
$a^2/2$.

The specular reflectivity was defined as
$R_s(k)=|r_m(k)|^2-w_1(k)$, because scattering decreases specular
reflectivity. For the total reflectivity we used the expression
\begin{equation}\label{chi838}
R(k)=\int\limits_{k-\delta}^{k+\delta}R_s(k)\fr{dp}{2\delta}+n_b+w_1(k),
\end{equation}
which contains two more fitting parameters: $\delta$ and $n_b$.
Thus in total we had 5 fitting parameters. The result of fitting
is shown in Figure~\ref{fit5}. For upper limit of the integral
$b=9$ in (\ref{chi883}) the $\chi^2$ was 8.5. Though it is too
high, it decreased considerably from value 31. The most important
result is that fitting shows a dip in reflectivity near the edge.
Fitting parameters were found to be: $c=8.3\cdot10^{-4}$, which
corresponds to randomness of positions of all the atoms in the
glass; $a^2=0.204$, which means that correlation length is near 40
\AA; resolution $\delta=2.2$\%, which is better than in previous
fittings; background $n_b=2.4\cdot10^{-3}$ is nearly the same as
before; and $2h^2=0.048$, i.e. $h\approx15$\AA, which means that
in preparation of the float glass some Sn atoms diffused into the
glass to the depth of 15\AA. Results of fitting shows that the dip
near the edge is quite well described by nonspecular reflectivity,
and is not a result of some instrumental peculiarity. The dip in
$\chi$ near the potential edge means that measured quantity is
larger than theoretical one. It happens because in theory we does
not take into account the neutrons scattered into the glass. These
scattered neutrons after going through the sample can be
registered by the detector. Because of them (the scattering is
maximal near the edge due to Yoneda effect) experimental result is
higher than theoretical one.

\subsection{Scattering on roughnesses at the
interface.} The usual approach is to consider roughness as a
Gaussian process. It means that all the rough surface is treated
as a single inhomogeneity and the scattered wave function is put
down as
\begin{equation}\label{agn}
\Psi_{s}(\kk,\rr)=\int d^2p_\|\exp(i\p\rr)A(\kk,\p),
\end{equation}
where
\begin{equation}
A(\kk,\p)=-N_0b\frac{it_{0s}(p_\bot)t_{0s}(k_\bot)}{2\pi
p_\bot}f(\ka'). \label{mrvm99}
\end{equation}
Here
\begin{equation}\label{gapr}
f(\ka)=\int d^2r'_\|\int\limits_0^{\zeta(\rr'_\|)}
dz'\exp(i\ka\rr')=\int
\fr{d^2r'}{i\kappa_\bot}\exp(i\ka_\|\rr'_\|)[\exp(i\kappa_\bot
\zeta(\rr'_\|))-1]\Theta(\zeta(\rr'_\|)>0),
\end{equation}
if roughness is a cavity (note that its scattering density is
$-N_0b$, and $\ka'=(\ka_\|,k'_\bot+p'_\bot)$, where
$q'_\bot=\sqrt{q^2_\bot-u_0}$), and
$$A(\kk,\p)=N_0b\int d^2r'_\|\exp(i\ka_\|\rr'_\|)\int\limits_{\zeta(\rr'_\|)}^0
dz'\Theta(\zeta(\rr'_\|)<0)\times$$
\begin{equation}
\times[\exp(ik_\bot z')+r_{s0}(k_\bot)\exp(-ik_\bot
z')][\exp(ip_\bot z')+r_{s0}(p_\bot)\exp(-ip_\bot z')],
\label{mrvm97}
\end{equation}
if roughness is a bump above\footnote{In our geometry, where
medium is at $z>0$ the bump is below the interface.} the average
interface. The parameter $\zeta$ is a random variable with
probability density distribution
\begin{equation}\label{gapr1}
P(\zeta)=\fr1{\sigma\sqrt{2\pi}}\exp(-\zeta^2/2\sigma^2),
\end{equation}
where $\sigma$ characterizes the average height of roughnesses.
Averaging both of expressions over $\zeta$ we obtain corrections
to specular reflectivity amplitude. It will be a combination of
error function $\Phi(q\sigma)$, where $q=k_\bot\pm p_\bot$ or
$q=k'_\bot+p'_\bot$.

For averaging of $|A(\kk,\p)|^2$, which depends on random
variables $\zeta$ at two different points $\rr'_\|$ we need the
density distribution of the Gaussian process
\begin{equation}\label{prden}
P(\zeta_1,\zeta_2)=\fr{1}{2\pi\sigma^2\sqrt{1-K^2(\rr_\|)}}
\exp\left(-\fr{\zeta_1^2+\zeta_2^2-2\zeta_1\zeta_2K(\rr_\|)}{2\sigma^2(1-K^2(\rr_\|))}\right),
\end{equation}
where $K$ is a correlation function, which depends on distance
$r_\|$ between two points, where $\zeta_{1,2}$ are defined.

\paragraph{Small roughnesses}

Though calculations with these formulas can be done up to the end
without principal difficulties, we shall not proceed this
complicated way and simplify our task assuming that the height,
$\sigma$, of roughnesses is sufficiently small~\cite{stey}, i.e.
$\sigma k_\bot\ll1$. For small grazing angles it means that
$\sigma\ll100$ \AA, which is quite practical. In that case we can
accept $A(\kk,\p)$ in the form
\begin{equation}
A(\kk,\p)=-N_0b\frac{it_{0s}(p_\bot)t_{0s}(k_\bot)}{2\pi
k_\bot}\int d^2r'_\|\zeta(\rr'_\|)\exp(i\ka\rr') \label{mrvm96}
\end{equation}
for all positive and negative $\zeta$. With this function we do
not have corrections to $\ora{r_0}$, because
$\langle\zeta\rangle=0$. The scattered waves are determined by
\begin{equation}\label{mr98}
\int d^2r'_{1\|}\exp(i\ka_\|\rr'_{1\|})\int
d^2r'_{2\|}\exp(i\ka_\|\rr'_{2\|})\zeta(\rr'_{1\|})\zeta(\rr'_{2\|}).
\end{equation}
Averaging of $\zeta(\rr'_{1\|})\zeta(\rr'_{2\|})$ over
(\ref{prden}) gives
\begin{equation}\label{prden2}
\langle\zeta(\rr'_{1\|})\zeta(\rr'_{2\|})\rangle=\sigma^2K(\rr'_{1\|}-\rr'_{2\|}).
\end{equation}
Therefore
\begin{equation}\label{prden3}
\langle|A(\kk,\p)|^2\rangle=S\fr{|N_0b\sigma|^2}{(2\pi p_\bot)^2
}\left|t_{0s}(p_\bot)t_{0s}(k_\bot)\right|^2\int
d^2r'_\|\exp(i\ka_\|\rr'_{\|})K(r'_\|),
\end{equation}
and indicatrix of nonspecular scattering is
\begin{equation}\label{prden4}
w(k\to
p)=2\pi\fr{|N_0b\sigma|^2}{k}\left|t_{0s}(p)t_{0s}(k_\bot)\right|^2
\int d^2r'_\|\exp(i\ka_\|\rr'_{\|})K(r'_\|).
\end{equation}
To finish calculations we need to define correlation function. It
is natural to suppose that
\begin{equation}\label{prden5}
K(\rr_\|)=\exp(-r_\|^2/2l^2),
\end{equation}
where $l$ is correlation length, or average dimension of
roughnesses along the interface. With this function we have
\begin{equation}\label{prden5a}
K_F(\ka_\|)=\int
d^2r'_\|\exp(i\ka_\|\rr'_{\|})\exp(-r_\|^2/2l^2)=2\pi
l^2\exp(-l^2\ka_\|^2/2),
\end{equation}
and substitution into (\ref{prden4}) gives
\begin{equation}\label{prden6}
w(k\to
p)=|N_0bl\sigma|^2\fr{(2\pi)^2}{k}\left|t_{0s}(p)t_{0s}(k_\bot)\right|^2
\exp(-l^2\ka_\|^2/2).
\end{equation}

\subsection{Scattering and fitting of periodic chains}

There are a lot of opportunities how to include roughness at
interfaces in multilayer systems. We can suppose that roughnesses
are independent on every interface, or they can correlate between
interfaces. It seems, that with sufficiently many fitting
parameters it is possible to fit any result. However we shall not
go this way. Because of so many opportunities it is better first
to study experimentally the angular distribution of non specularly
reflected neutrons, find its distinctive features and after that
compare them with theoretical predictions based on different
theoretical models. This is the way we are going proceed further.

\section{Supermirror}

\begin{figure}[t!]
{\par\centering\resizebox*{12cm}{!}{\includegraphics{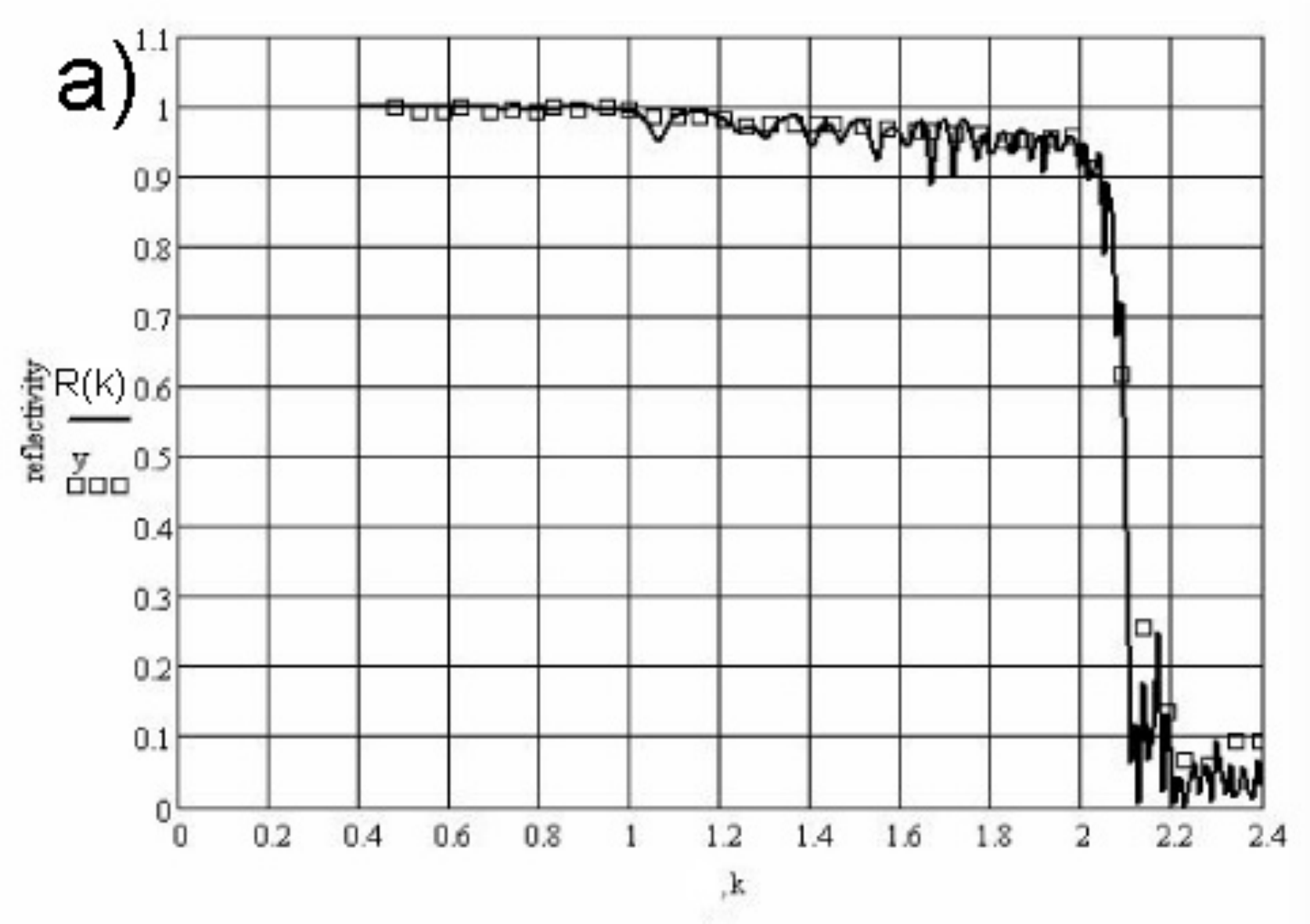}
\hspace{15mm}\includegraphics{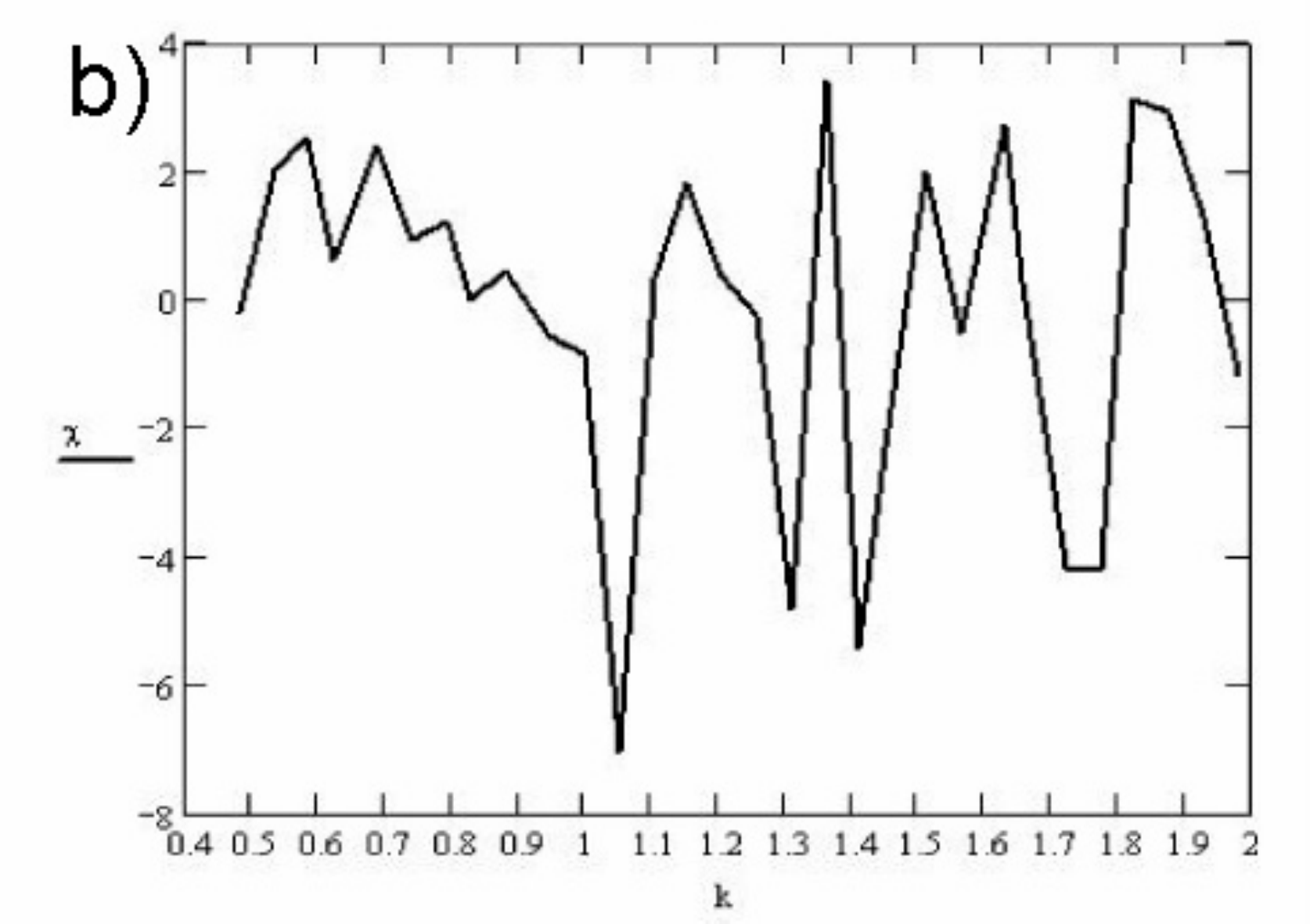}}
\par}
\caption{\label{sm2}a)Experimental data points and theoretical
calculation of supermirror M2.b) Distribution $\chi(k)$. It is
obtained from direct comparison of experimental and theoretical
date without fitting.}
\end{figure}

Besides of the samples described above there was also prepared in
BNC Budapest a supermirror M2, which consisted of 8 periodic
chains and total number of 59 bilayers. The periods and number of
them were found according to the above prescription with some
corrections. The result of measurements comparing to calculations,
in which we put Ni potential to be 0.964-0.005i and Ti potential
-0.26-0.005i, is shown in Figure~\ref{sm2}a). We see good
coincidence. If we limit calculation of $\chi^2$ to the range
$k<2$, then $\chi^2=7$, which is not bad, if to take into account
that there were no fitting at all except some guess about
imaginary parts of the potentials. The imaginary parts are higher
than table ones because of possible impurities, inhomogeneities
and surface roughness.

\section{Conclusion}

Cooperation of theoreticians and experimentalists in research of
multilayer systems is found to be very fruitful. We see that
technology of preparation of such systems by Mirrotron Ltd,
Budapest is good, but it can be further improved after analysis of
surface imperfection and their correlation with parameters of
producing systems. This analysis can be performed with new
experiments aimed at investigation of diffuse scattering and
angular distribution of reflected neutron with better angular
resolution.

\section*{Acknowledgement}
One of us V.K.I. is grateful to Yu.V.Nikitenko for support.

\end{document}